\documentclass[12pt]{article}

\usepackage{url}
\usepackage{setspace}
\usepackage{amsmath,amssymb,amscd, amsthm}
\usepackage{bbold}
\usepackage{amsfonts}
\usepackage[usenames,dvipsnames]{color}
\usepackage{graphicx}
\usepackage{subcaption}
\usepackage{cite}

\usepackage{hyperref}

\def\zb{\overline{z}}
\def\wb{\overline{w}}

\usepackage{cleveref}

\usepackage[width=0.8\textwidth]{caption}

\renewcommand{\inf}{{\infty}}
\newcommand{\calL}{\mathcal{L}}
\newcommand{\sltwo}{$\lie{sl}(2,\RR)$ }

\def\Lc{\mathcal{L}}
\def\Lcb{\overline{\mathcal{L}}}
\def\p{\partial}
 \newcommand{\bea}{\begin{eqnarray}}
\newcommand{\eea}{\end{eqnarray}}
\newcommand{\ba}{\begin{align}}
\newcommand{\ea}{\end{align}}

\newcommand{\CC}{\mathbb{C}}
\newcommand{\RR}{\mathbb{R}}

\newcommand\rref[1]{(\ref{#1})}

\newcommand{\Fc}{\mathcal{F}}
\newcommand{\ie}{{\it i.e.~}}

\newcommand{\lie}[1]{\mathfrak{#1}}
\DeclareMathOperator{\Tr}{Tr}
\newcommand{\intertwine}[3]{\mathcal{I}_{#1;#2,#3}}
\newcommand{\lw}[1]{|\text{LW}_{#1}\rangle}
\newcommand{\dw}{2\pi\tau}

\begin{document}

\title{\vspace{-1cm}
	\begin{flushright}\end{flushright}
	\bf{Witten Diagrams for Torus Conformal Blocks}
	\vspace{12pt}}
\author{
	Per Kraus,$^{1}$
	Alexander Maloney,$^{2}$
	~Henry Maxfield$^{2}$, \\ Gim Seng Ng$^{2}$, and Jie-qiang Wu$^{3}$
	\\ \\
	$^{1}$ Mani L. Bhaumik Institute for Theoretical Physics
	\\Department of Physics and Astronomy, University of California
	\\
	Los Angeles, CA 90095, USA
	\\
	\\
	$^{2}$ Physics Department, McGill University
	\\
	Montr\'eal, QC H3A 2T8, Canada
	\\
	\\
	$^{3}$ Physics Department, Peking University,
	\\ Beijing 100871, China
}
\maketitle

\vspace{-1em}

\abstract{
We give a holographic description of global conformal blocks in two dimensional conformal field theory
on the sphere and
on the torus.  We show that the conformal blocks for one-point functions on the torus can be written as Witten diagrams in thermal AdS. This is accomplished by deriving a general conformal Casimir equation for global conformal blocks, and showing that Witten diagrams obey the same equation. We study the semi-classical limit of $n$-point conformal blocks, and show that these equal the action of a network of bulk world-lines obeying appropriate geodesic equations.  We give an alternate description in the Chern-Simons formulation of 3D gravity, where the conformal blocks are described by networks of Wilson lines, and argue that these formulations are equivalent.}

\clearpage

\tableofcontents

\section{Introduction}

Recent years have seen remarkable progress in the study of conformal field theory, where unitarity and symmetry can be used to constrain the dynamics without the need for a perturbative expansion (see \cite{Rychkov:2016iqz,Poland:2016chs,Simmons-Duffin:2016gjk,Penedones:2016voo} for a review of recent progress).
This may potentially lead to a general understanding of the emergence of AdS gravitational physics from conformal field theory.
An important development in this subject has been the use of AdS space as a tool for organizing CFT kinematics.
In this paper we focus on two dimensional CFTs, where symmetry constraints are the strongest, although many results can be generalized to higher dimensions.
We will study the bulk gravitational interpretation of global conformal blocks at zero and finite temperature, where they can be represented as Witten diagrams in the appropriate gravitational background.

The states and operators of a conformal field theory can be organized into representations of conformal symmetry.
A conformal block is the contribution of a particular representation (or representations) to a given physical observable.  A block is a purely kinematic object, in that it is uniquely determined by the symmetry structure of the theory and the choice of representation(s).
The conformal symmetries of a CFT$_d$ are precisely the isometries of $AdS_{d+1}$, so it is natural to expect that any conformal block can be rewritten in the language of quantum fields in AdS.  This program was carried out explicitly in \cite{Hijano:2015zsa} where the conformal blocks for CFT four-point functions of scalar operators were rewritten in terms of AdS geodesic Witten diagrams.  Further work in this direction, including the presence of external operators with spin, may be found in \cite{Bhatta:2016hpz,Besken:2016ooo,Czech:2016xec,Nishida:2016vds,Castro:2017hpx,Dyer:2017zef,
Sleight:2017fpc,Chen:2017yia,Gubser:2017tsi}.  The Virasoro symmetry present for $d=2$ allows one to define Virasoro blocks in this case, and bulk representations for Virasoro blocks, mainly in the so-called heavy-light limit, can be found in \cite{Fitzpatrick:2014vua,Hijano:2015rla,Alkalaev:2015wia,Hijano:2015qja,Alkalaev:2015lca,Alkalaev:2015fbw,
Guica:2016pid,Alkalaev:2016rjl}.
The advantage of this approach is that a-priori complicated CFT objects (which sometimes cannot be computed explicitly) can often be given an extremely simple bulk interpretation.  Thus AdS appears as a useful tool for organizing CFT observables.

In this  paper our primary interest is the bulk interpretation of finite temperature global conformal blocks in two dimensional CFT, i.e. the conformal blocks for correlation functions on the torus.
This is particular interesting because at high temperature the bulk dual is an AdS black hole, so the holographic description can be interpreted in terms of bulk dynamics in a black hole background.

As a simple example, let us consider the case of a one-point function of an operator $O_1$ at finite temperature:
\begin{equation}\label{sauron}
\langle O_1 \rangle (\beta) = \sum_i \langle i | O_1 | i \rangle e^{-\beta E_i} = \sum_{\alpha \text{ primary}} \langle \alpha | O_1 | \alpha \rangle~
\left |\mathcal{F}(h_\alpha, h_1, \beta)\right|^2
\end{equation}
Here we have expanded in terms of conformal blocks\footnote{We write $|\mathcal{F}|^2$ for the conformal blocks since in two dimensions they factorize into holomorphic and antiholomorphic parts $\mathcal{F}\bar{\mathcal{F}}$, which we will have occasion to consider separately.}
$|\mathcal{F}(h_\alpha, h_1, \beta)|^2 =  e^{-\beta (h_\alpha+{\bar h}_\alpha - c/12)} \left(1 + \dots\right)$
which describe the contribution from all descendant operators built on top of a primary operator $\alpha$.  Our question is simple: what is the bulk AdS interpretation of the expansion \rref{sauron}? Let us imagine working at low temperature, and consider the contribution from a single primary operator $\alpha$.  This operator is dual to a bulk field in AdS, with mass and spin determined by the dimensions $(h_\alpha, {\bar h}_\alpha)$.  The factor $e^{-\beta (h_{\alpha} + {\bar h}_\alpha-c/12)}$ is the Boltzmann factor for this particle sitting at the origin.  In Euclidean signature, this is the action of a bulk worldline at the origin of AdS wrapping the thermal circle.

\begin{figure}
 \centering
  \includegraphics[width=0.6\textwidth]{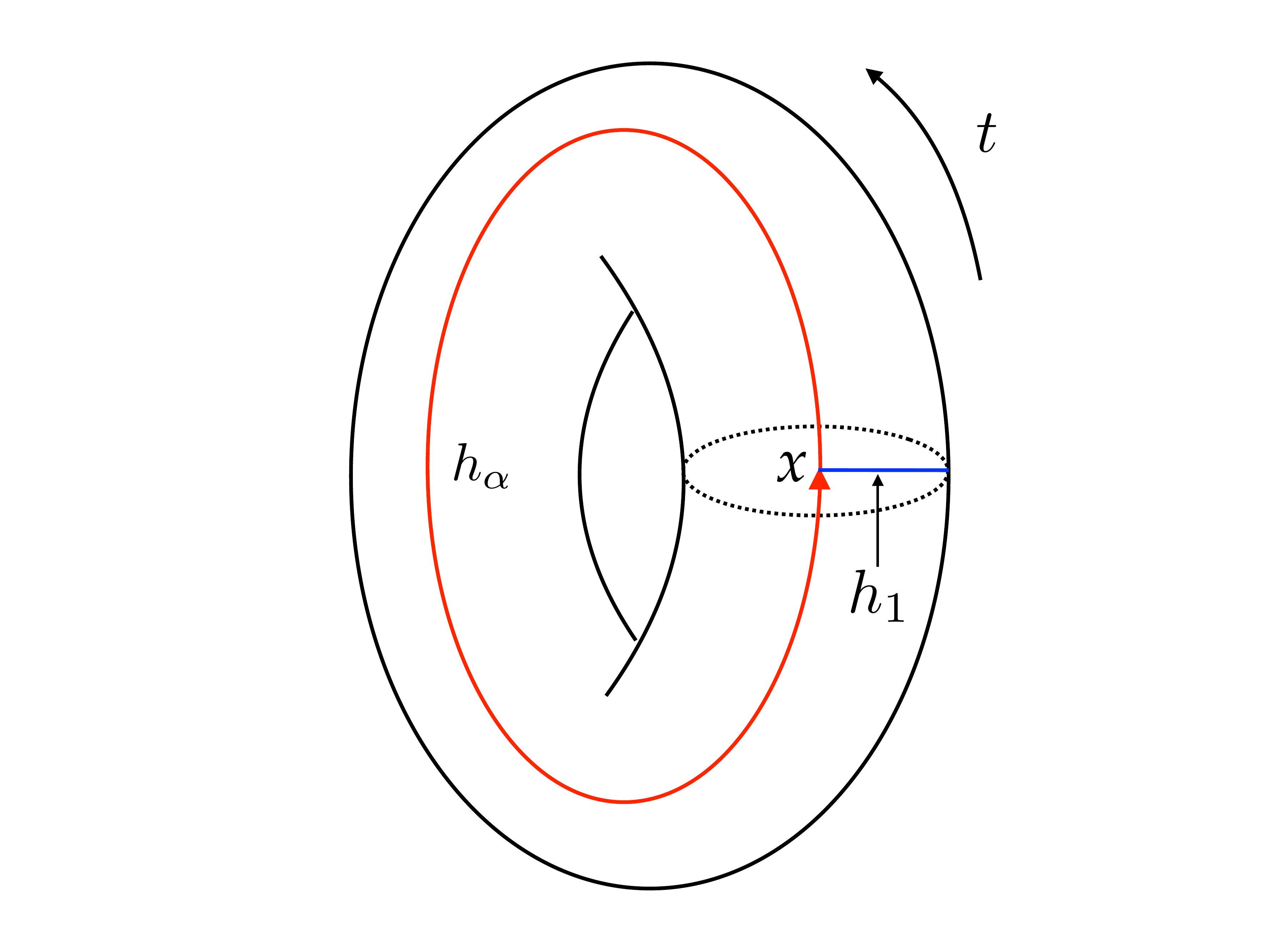}
 \caption{Bulk dual of the torus one-point block.
 The blue line represents the bulk-to-boundary propagator for $h_1$, and the red line the propagator for the $\alpha$ particle to propagate once around the thermal circle.  \label{fig:block}
}
 \end{figure}

We will show that $|\mathcal{F}(h_\alpha, h_1, \beta)|^2$ is precisely equal to the bulk Witten diagram for a particle which propagates once around the thermal circle, as in Fig.~\ref{fig:block},
\begin{equation}\label{beorn}
|\mathcal{F}(h_\alpha, h_1; \beta)|^2 = \int\displaylimits_\text{Thermal AdS} \mkern-20mu d^3x \sqrt{g} G^{(h_1)}_{b\partial} (x,w) G_1^{(h_\alpha)}(x,\beta)~.
\end{equation}
Here $G^{(h_1)}_{b\partial} (x,w)$ is the bulk-boundary propagator for the $O_1$ particle to propagate from a point $x$ in  thermal AdS to a point $w$ on the boundary.  On the other hand, $G_1^{(h_\alpha)}(x,\beta)$ is the bulk-bulk propagator which describes the $\alpha$ particle propagating exactly once around the thermal circle.
This differs from the full bulk-bulk thermal propagator, which would involve a sum over windings around the thermal circle, by terms which are exponentially suppressed at low temperature.
Equation \rref{beorn} should be viewed as the generalization of the results of \cite{Hijano:2015zsa} to thermal blocks.
This means that the sum \rref{sauron} can be interpreted as a sum over particles propagating in a thermal AdS background.
In this paper we will consider only global blocks, which in the bulk language means that the particle does not back-react on the geometry.  We expect that the full Virasoro block should account for gravitational back-reaction.

 This result can also be interpreted in terms of particles propagating in the BTZ black hole geometry, following \cite{Kraus:2016nwo}.  In particular, since correlation functions on the torus are modular covariant, we can write the one point function as
\begin{equation}\label{saruman}
\langle O_1 \rangle (\beta) =
\beta^{-(h_1 + {\bar h}_1)}
\sum_{\alpha \text{ primary}} \langle \alpha | O_1 | \alpha \rangle~
\left |\mathcal{F}\left(h_\alpha, h_1, {4\pi^2 \over \beta}\right)\right|^2~.
\end{equation}
Here the sum is over states propagating around the spatial circle on the torus, rather than a sum over states propagating around the thermal circle.  In the bulk, $\left|\mathcal{F}\left(h_\alpha, h_1, {4\pi^2 \over \beta}\right)\right|^2$ is now interpreted as a Witten diagram in the BTZ black hole background.
The advantage of this approach is that the contribution to \rref{saruman} from light states will now dominate the behaviour of $\langle O_1\rangle(\beta)$ at high temperature.
For example, the lightest state $\alpha$ with non-vanishing one point function gives the leading asymptotics
\begin{equation}\label{isengard}
\langle O_1\rangle(\beta) \sim \beta^{-(h_1 + {\bar h}_1)}
\langle \alpha | O_1 | \alpha\rangle
\exp\left\{-4\pi^2 (h_{\alpha} + {\bar h}_\alpha-c/12)/\beta\right\}+\cdots
\end{equation}
Indeed, in \cite{Kraus:2016nwo} it was argued that the Witten diagram for an $\alpha$ particle which wraps the horizon once will reduce precisely to \rref{isengard} in the high temperature limit.  In this limit the $\alpha$ particle just sits at the event horizon.
We can now understand the subleading corrections to \rref{isengard}, as captured by the full sum \rref{saruman}.  In particular, we see that if we include the full tower of descendant states built on top of $\alpha$, this simply describes the propagation $\alpha$ particle in the Euclidean BTZ background, rather than just sitting at the horizon.

In fact, we will see that many of these results can be generalized from one-point functions to $n-$point functions.  In order to do this, we will introduce a new method for the computation of sphere and torus blocks in two dimensional CFT.  Rather than a direct computation (as in \cite{Hadasz:2009db} for the case of the torus one-point block), we will instead derive a general conformal Casimir equation which is obeyed by these conformal blocks, similar to that obeyed by sphere four-point blocks.  We will work out the case of the one-point block in detail, and show that this description leads immediately to the bulk description in terms of the Witten diagram described above.  We will generalize this to $n-$point functions, where it is difficult to find explicit expressions for the blocks, and discuss the bulk Witten diagram description of $n-$point conformal blocks on the sphere and on the torus.

We will then move on to study the semi-classical limit, where a bulk Witten diagram can be approximated by the action of a collection of bulk geodesics.  For example, in this limit the one-point block is computed by the action of a pair of bulk geodesics -- the blue line and the red circle in Figure 1 -- one of which wraps the thermal circle.  The dynamics of this pair of geodesics is still somewhat complicated, since the geodesics will pull on one another in a non-trivial way (which depends on the particle masses) to reach a configuration that minimizes the total worldline action.\footnote{Related considerations were discussed in \cite{Alkalaev:2016ptm, Alkalaev:2016fok}.}
We will show that the Casimir equations for our $n$-point conformal blocks reduce precisely to the correct equations of motion for these geodesics.

We will conclude by giving an alternate description of these results in the language of Chern-Simons theory.  In this case we show that the conformal blocks can be computed in terms of a network of bulk Wilson lines, following \cite{Bhatta:2016hpz,Besken:2016ooo,Fitzpatrick:2016mtp}.  In this description, the thermal blocks are now evaluated in a Chern-Simons background which has non-trivial holonomy around the thermal circle.

\section{Torus one-point block from a Casimir equation}
\label{sec:torusonepointCasimir}

In this section, we derive the global conformal block for torus one-point functions using a Casimir equation. The result  is known in the literature \cite{Hadasz:2009db}, but the derivation using the Casimir equation is new and will be useful later for the holographic computations as well as multi-point blocks.

The (holomorphic) torus 1-point block is defined as
\begin{equation}\label{blockdef}
\mathcal{F}(h_{\alpha},h_1;q)=\Tr \left[ P_{\alpha}q^{L_0}\phi_1(w)\right] \,.
\end{equation}
Here, the ``external" operator $\phi_1(w)$ is a quasi-primary field on the cylinder, with coordinate $w$. The generators of the holomorphic \sltwo are $L_0,L_{\pm 1}$, satisfying the commutation relations $[L_m,L_n] = (m-n)L_{m+n}$, with $L_0$ generating translations in $w$. The insertion of the projection operator $P_\alpha$ has the effect of restricting the trace to a sum over states built on the quasiprimary state $|h_\alpha\rangle$. Up to an overall factor of  $\langle h_\alpha|\phi_1(0)|h_\alpha\rangle$, which is the three-point coupling $\phi_\alpha \phi_\alpha \phi_1$, the one-point block is fully determined by \sltwo symmetry. We shall drop this overall factor for the rest of the paper.
\begin{figure}
\centering
  \includegraphics[width=4cm]{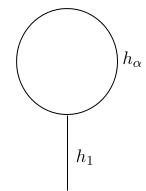}\\
  \caption{1-point block on torus}\label{1pttorus}
\end{figure}
 The one-point block is depicted in Fig.~\rref{1pttorus}, with the ``internal" operator $\phi_\alpha$ circulating in the loop and fusing with the external operators $\phi_1$ at a three-point vertex.

It is straightforward enough to compute the one-point block directly: simply enumerate the states contributing to the trace, and use the  \sltwo  commutation relations to compute the matrix elements, as was done in \cite{Hadasz:2009db}. However, as is the case for four-point blocks on the sphere, it is much more elegant and illuminating to proceed by using the Casimir operator to derive a second order differential equation obeyed by the block.

We proceed by inserting the quadratic Casimir operator
\begin{equation}
\label{Casimira}
 L^2=\frac{1}{2}(L_1L_{-1}+L_{-1}L_1)-L_0^2~,
\end{equation}
into the definition of the block, \ie we want to compute $\Tr \left[L^2 P_{\alpha}q^{L_0}\phi_1(w)\right]$. The Casimir commutes with the algebra, so $L^2$ is constant on the representation, with eigenvalue $-h_\alpha(h_\alpha-1)$, from which we immediately conclude that
\begin{equation}
\label{Casimirb}
 \Tr \left[L^2 P_{\alpha}q^{L_0}\phi_1(w)\right] = -h_\alpha(h_\alpha-1)\mathcal{F}(h_{\alpha},h_1;q)\,.
\end{equation}
On the other hand, we can also derive some identities by commuting the generators in $L^2$ through the other operators and using cyclicity of the trace.  For example, %
\begin{align}
	 \Tr\left[P_\alpha L_n  q^{L_0} \phi_1(w)\right] &= q^n\Tr\left[ P_\alpha q^{L_0}L_n \phi_1(w)\right]\nonumber\\
&= q^n\Tr \left[ P_\alpha q^{L_0} \phi_1(w)L_n\right]  +q^n\Tr\left[P_\alpha q^{L_0}[L_n ,\phi_1(w)]\right]\nonumber\\
&=  q^n\Tr \left[ P_\alpha q^{L_0} \phi_1(w)L_n\right]  +q^n\Lc_n\Tr \left[ P_\alpha q^{L_0}\phi_1(w)\right] \,,
\end{align}
where we have also used the action of the symmetry generators on a quasiprimary field by the differential operator $\Lc_n$
\begin{equation}
\label{eq:cylLn}
 [L_n,\phi(w)] =-\Lc_n \phi(w)\quad \text{with}\quad  \Lc_n \phi(w)=e^{-inw} \big( nh +i \p_w\big)\phi(w)\quad (n=-1,0,1)
\end{equation}
and the identity
\begin{equation}\label{eq:useful}
L_n q^{L_0} = q^{L_0+n} L_n \,.
\end{equation}
 Thus, we have shown that for $n=\pm 1$, we have
\begin{equation}
 \Tr \big[P_\alpha L_n q^{L_0} \phi_1(w)\big] = {q^n \over 1-q^n} \Lc_n  \Tr \big[P_\alpha q^{L_0}\phi_1(w)\big]
 = {q^n \over 1-q^n} \Lc_n \mathcal{F}(h_{\alpha},h_1;q)
 \,.
\end{equation}
We may additionally insert $L_{-n}$ into the above relation and repeat the same steps to obtain
\begin{equation}
 \Tr \big[P_\alpha L_{-n} L_n q^{L_0} \phi_1(w)\big] =  {2nq^n \over 1-q^n}q\p_q  \mathcal{F}(h_{\alpha},h_1;q)
 +{1\over (1-q^n)(1-q^{-n})}\Lc_n\Lc_{-n}
  \mathcal{F}(h_{\alpha},h_1;q)\,.
\end{equation}
Insertions of $L_0$ can be accounted for by taking derivatives with respect to $q$:
\begin{equation}
 \Tr \left[
 P_\alpha L_0^2 q^{L_0} \phi_1(w)
 \right] =q\p_q \left( q \p_q
   \mathcal{F}(h_{\alpha},h_1;q)
 \right)\,
\end{equation}
Combining these results, we rewrite the insertion of the Casimir as a differential operator (in $q$) acting on the block:
\bea
 \Tr \big[P_\alpha L^2 q^{L_0} \phi_1(w)\big]
 = -\left[ q\p_q  q \p_q  -{1+q\over 1-q}q\p_q   +{q\over (1-q)^2} {1\over 2}\{ \Lc_1,\Lc_{-1}\}\right]
    \mathcal{F}(h_{\alpha},h_1;q)
    \,,
\eea where $\{ \Lc_1,\Lc_{-1}\} \equiv \Lc_1 \Lc_{-1}  + \Lc_{-1} \Lc_{+1}$.
Since $\Lc_0$ annihilates $\Tr\big[ P_\alpha q^{L_0} \phi_1(w)\big]$ by translation invariance (\ie $\mathcal{F}(h_{\alpha},h_1;q)
   $ is in fact independent of $w$), this can be written as
\begin{equation}
\Tr \big[P_\alpha L^2 q^{L_0} \phi_1(w)\big]
= -\left[ q\p_q  q \p_q  -{1+q\over 1-q}q\p_q   +{q\over (1-q)^2} \Lc^2\right]
    \mathcal{F}(h_{\alpha},h_1;q)\,.
\end{equation}
Now, using Eq.~(\ref{Casimirb}) and $\Lc^2 \phi_1(w)=-h_1(h_1-1)\phi_1(w)$,  we arrive at
\begin{equation}
-h_\alpha(h_\alpha-1)   \mathcal{F}(h_{\alpha},h_1;q)
  = -\left[ q\p_q  q \p_q  -{1+q\over 1-q}q\p_q   -{q\over (1-q)^2} h_1(h_1-1)\right]
      \mathcal{F}(h_{\alpha},h_1;q)\,,
\end{equation}
which can be rewritten as
\begin{equation}\label{diffeq}
\Big[  q(1-q)^2\p_q^2 -2q(1-q)\p_q -h_\alpha(h_\alpha-1)q^{-1}(1-q)^2 -h_1(h_1-1)\Big]  \mathcal{F}(h_{\alpha},h_1;q)=0\,.
\end{equation}
This is essentially a hypergeometric equation, and the  solution with the correct small $q$ asymptotics
(\ie the solution which behaves as $q^{h_{\alpha}}$ as $q\rightarrow 0$)  is
\begin{align}
    \mathcal{F}(h_{\alpha},h_1;q) &= \frac{q^{h_\alpha} }{ (1-q)^{h_1}} {}_2F_1(1-h_1,2h_\alpha-h_1;2h_\alpha;q) \nonumber \\
    &=\frac{ q^{h_\alpha} }{ (1-q)^{1-h_1} } {}_2F_1(h_1,2h_\alpha+h_1-1,2h_\alpha,q)\,.
\end{align}
Notice that if we set $h_1=0$ we get $ \mathcal{F}(h_{\alpha},0;q)= q^{h_{\alpha}}/ (1-q)$ which is the \sltwo  character of the representation built on $|h_\alpha\rangle$. Furthermore, from the Casimir method or explicitly from the above solution, it is obvious that the block with external operator with dimension $h_1$ is the same as its ``shadow''
which has dimension $1-h_1$, \ie $ \mathcal{F}(h_{\alpha},h_1;q)= \mathcal{F}(h_{\alpha},1-h_1;q)$.

For future reference we define the differential operator
\begin{equation}\label{Qdef} Q_h = q(1-q)^2\p_q^2 -2q(1-q)\p_q -h(h-1){(1-q)^2\over q}
\end{equation}
so that Eq.~(\ref{diffeq}) now reads like an eigenvector equation:
\begin{equation}\label{onepointeq} Q_{h_{\alpha}}  \mathcal{F}(h_{\alpha},h_1;q) = h_1(h_1-1)  \mathcal{F}(h_{\alpha},h_1;q)\,
\end{equation}

\section{Holographic description of the torus one-point blocks}

In this section, we will describe the bulk dual of a torus one-point block.  We will begin with a general proposal for the bulk representation of the conformal block, before proving a bulk-bulk propagator identity which will imply that our bulk proposal satisfies the same Casimir equation as the boundary torus one-point block.

\subsection{Generalities}

We work with the global AdS$_3$ metric in the form
\begin{equation} ds^2 = {1\over\cos^2 \rho}(d\rho^2 + dt^2 +\sin^2 \rho \; d\phi^2)\end{equation}
and define the complex coordinate $w=\phi+it$. The AdS metric has isometry group $\lie{sl}(2,\RR)\times\lie{sl}(2,\RR)$. In this section, $L_n$ and $\Lc_n$ will denote these generators acting in the bulk.   In particular, the isometry generators are $\Lc_{\pm 1, 0}$ and $\Lcb_{\pm 1,0}$, which obey the \sltwo algebra
\begin{equation} [\Lc_m,\Lc_n] = (m-n)\Lc_{m+n}\;,\quad  [\Lcb_m,\Lcb_n] = (m-n)\Lcb_{m+n} \,.
 \end{equation}
The quadratic Casimir is the differential operator\footnote{Explicitly, the non-zero components of $\eta^{AB}$ are $\eta^{00}=-1$ and $\eta^{+1,-1}=\eta^{-1,+1}=1/2$.}
\begin{equation}
\label{eq:bulkCasimir}
\Lc^2 = \eta^{AB}\Lc_A \Lc_B
= \frac{1}{2}(\Lc_1 \Lc_{-1}+\Lc_{-1}\Lc_1)-\Lc_0^2 \; ; \quad  A=-1,0,+1
 \end{equation}
with eigenvalues $\Lc^2 = -h(h-1)$ when acting on the \sltwo representation built on $|h\rangle$ .  For a scalar primary (where $\bar{h}=h$), the relation to the scalar Laplacian on AdS is
\begin{equation}
 \label{eq:bulkCasimir2}
\Lc^2 = -{1\over 4}\nabla^2 \,.
\end{equation}
The bulk-bulk propagator $G_{bb}^{(h)}$ for a scalar field of mass $m^2=4h(h-1)$ obeys
\begin{equation}  \nabla^2 G_{bb}^{(h)}(x',x) =4h(h-1)G_{bb}^{(h)}(x',x)+{1\over \sqrt{g}}\delta^3(x'-x)
\end{equation}
The bulk-boundary propagator obeys the source free wave equation, with boundary condition $G^{(h)}_{b\p}(\rho,t,\phi; t',\phi') \sim (\cos \rho)^{2-2h} \delta^{(2)}(t,\phi; t',\phi')$.  The explicit forms of the propagators will not be needed.

Thermal AdS is obtained by making the identification $w \cong w+2\pi \tau$ for $\tau$ in the upper half-plane.  The bulk solution is then a solid torus, whose conformal boundary is a torus with modular parameter $\tau$.

\begin{figure}[h!]
 \centering
  \includegraphics[width=0.6\textwidth]{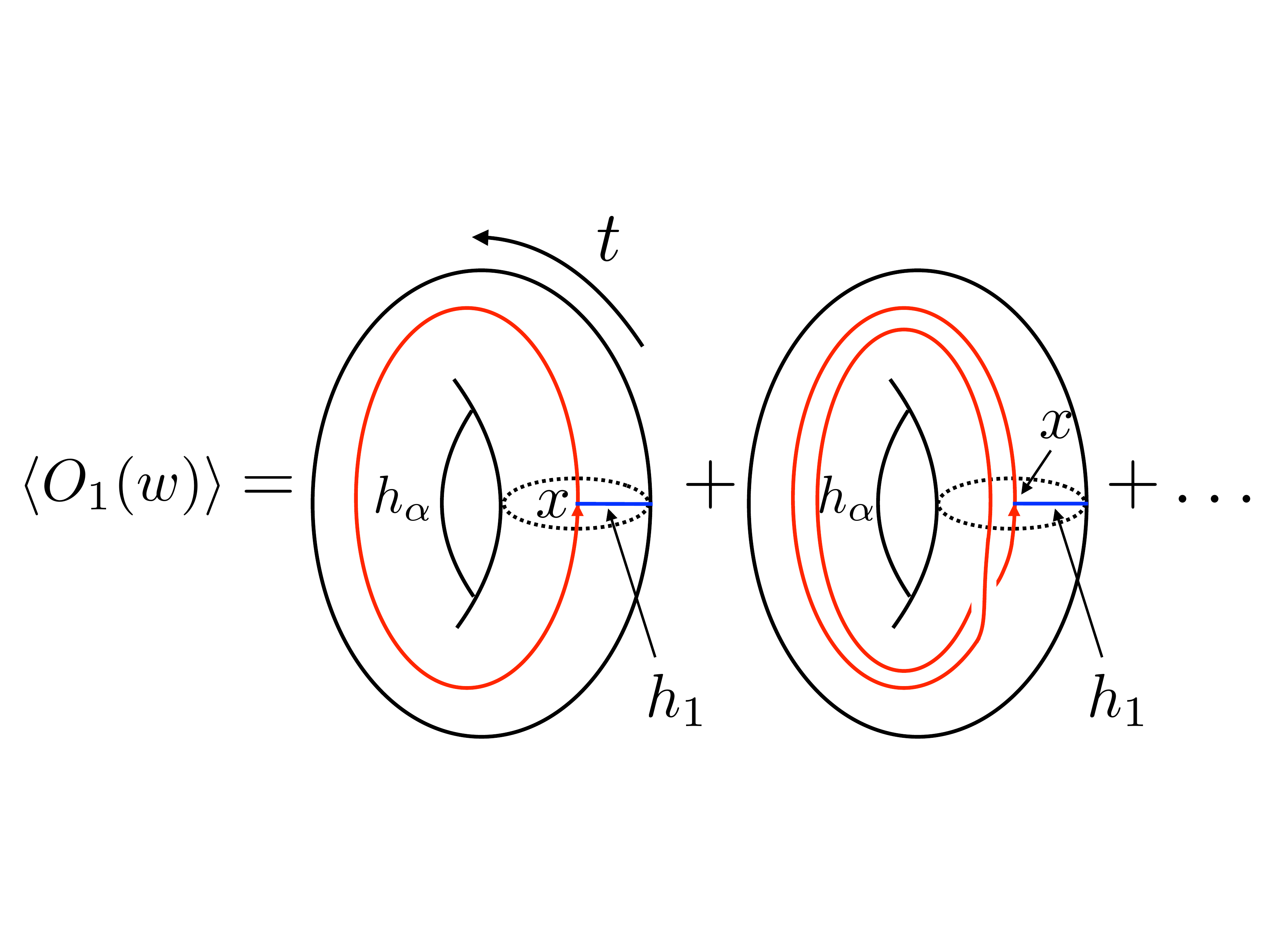}
 \caption{Bulk dual of the torus one-point function as a sum over bulk diagrams. The blue line represents the bulk-to-boundary propagator $G_{b\partial}^{(h_1)}(x)$. In the first diagram, the red line winding around the thermal circle once represents the $G_1^{(h_\alpha)}(x,q)$ contribution, while in the second diagram, the red line winding around the thermal circle twice represents the contribution from two windings around the thermal circle. The bulk point $x$ is to be integrated over all thermal AdS.
  \label{fig:HoloOnePtTwoWinding}
}
 \end{figure}

We now seek a bulk description the one-point torus conformal block $\Fc(h_\alpha,h_1;q)$ defined and computed in the last section.  This will involve introducing two fields in the bulk, namely scalars of mass $m_\alpha^2 =4h_\alpha(h_\alpha-1)$ and $m_1^2 = 4h_1(h_1-1)$.  These fields interact via the cubic coupling $\lambda \Phi_1 \Phi_\alpha^2$.    Now, given this setup, we can imagine computing the one-point Witten diagram $\langle O_1 \rangle$ to first order in $\lambda$,
\begin{equation} \langle O_1(w) \rangle =
\lambda \mkern-16mu \int\displaylimits_\text{Thermal AdS} \mkern-16mu d^3x \sqrt{g} G_{b\p}^{(h_1)}(x,w)G_{bb}^{(h_\alpha)}(x,x)~. \end{equation}
Here $x$ denotes a bulk point and $w$ a boundary point.
The propagators in thermal AdS can be obtained from those in global AdS by summing over images to respect the $w\cong w+2\pi \tau$ identification.  From a first-quantised worldline point of view, the sum over images of the bulk to bulk propagator  $G_{bb}^{(h_\alpha)}(x,x)$ is a sum over topologies of worldlines, organised by the number of windings around the thermal circle.
 Decomposing the contributions to $\langle O_1(w) \rangle $ according to their winding around the thermal circle yields the sum represented pictorially in Fig.~\ref{fig:HoloOnePtTwoWinding}.
  The zero winding contribution is divergent, but we omit this (equivalently we add a local counterterm to cancel it) since it corresponds to the one-point function in global AdS, which vanishes.

Now, the full Witten diagram is expected to be equal to a sum of one-point blocks, as  is familiar in the analogous case of four-point functions on the plane.  In the case at hand, the Witten diagram receives contributions from the infinite tower of multi-trace primary operators built out of products of the single trace primary $O_\alpha$ with insertions of derivatives. The question is how to isolate the contribution from a single block, in particular that of the single trace primary $O_\alpha$.

In the case of four-point blocks on the plane, part of the bulk prescription involved restricting the integration over interaction vertices to lie on bulk geodesics connecting the boundary operator insertion points.  This makes good intuitive sense, as it corresponds to computing ``part" of the full correlator, while respecting conformal invariance.  In the present case we have only a single boundary insertion so there is no natural geodesic over which to integrate the vertex.  A little thought reveals that the only natural thing to do is to isolate a single winding contribution in the full expression for the bulk-bulk propagator.  In particular, it seems natural to expect that the single winding terms yield the contribution from the single trace primary running in the loop, dual to the bulk one-particle states.   Similarly, we can expect the $n$-th winding sector yields contributions from primaries built out of the $n$-th power of $O_\alpha$, but note that there are many such primaries differing by insertions of derivatives, corresponding to the different possible wavefunctions of $n$-particle states in AdS, so for $n>1$ we get a sum over blocks rather than a single block.

Our proposal is therefore
\begin{equation}
\label{blkprop}  |\mathcal{F}(h_{\alpha},h_1;q)|^2 \sim  \int\displaylimits_\text{Thermal AdS} \mkern-16mu d^3x \sqrt{g} ~G_{b\p}^{(h_1)}(x,w)G_{1}^{(h_\alpha)}(x,q)~,
\end{equation}
where $G_{1}^{(h_\alpha)}(x,q)$ denotes the bulk-bulk propagator in global AdS with endpoints related by a single thermal translation.  On the other hand, $G_{b\p}^{(h_1)}(x,w)$ is the full bulk-boundary propagator in thermal AdS, obtained from the global AdS propagator by summing over all thermal translations.

In fact, there is an alternative representation of the proposal in \cref{blkprop}. This is given by
\begin{equation}
\label{blkprop2}  |\mathcal{F}(h_{\alpha},h_1;q)|^2 \sim  \int_\text{AdS} d^3x \sqrt{g} ~G_{b\p}^{AdS,(h_1)}(x,w)G_{1}^{(h_\alpha)}(x,q)~,
\end{equation}
Note that the integration in Eq.~(\ref{blkprop2}) is over all of global AdS while $G_{b\p}^{AdS,(h_1)}(x,w)$ is the bulk-boundary propagator on global AdS.
The equivalence between Eq.~(\ref{blkprop}) and Eq.~(\ref{blkprop2}) is apparent when one interprets Eq.~(\ref{blkprop}) as letting the interaction vertex go around the thermal circle any number of times, dragging with it the bulk-boundary propagator.\footnote{To see this equivalence more explicitly,
first note that $G_1^{(h_\alpha)}$ is independent of $t$ and $\phi$. Then, rewrite the full thermal AdS bulk-boundary propagator $G_{b\p}^{(h_1)}(x,w)$ as a thermal-image sum over the global AdS bulk-boundary propagator $G_{b\p}^{ (h_1)}(x,w)$. The thermal sum then converts the integration region from thermal AdS to global AdS.
} This is natural, as it corresponds to performing the integration over all configurations subject only to the constraint that the bulk-bulk propagator winds once around the thermal circle. However, as we will see later, the representation in  Eq.~(\ref{blkprop2}) will be more convenient for generalizations to higher-point torus blocks.

To prove Eq.~(\ref{blkprop}), in the next subsection, we will show that the RHS obeys the differential equation (\ref{onepointeq}) and also shares the same low temperature asymptotics as $\mathcal{F}(h_{\alpha},h_1;q)$. These two conditions uniquely fix $\mathcal{F}(h_{\alpha},h_1;q)$.

Before we go on, we should briefly mention convergence of the integral. Since the bulk-boundary propagator contains a non-normalizable delta-function supported piece, the integral over AdS has an IR divergence from the boundary near the point $w$ if $h_1>2h_\alpha$. To avoid this subtlety, we will restrict our considerations to the case $h_1<2h_\alpha$.

\subsection{A bulk-bulk propagator identity}

The conjecture Eq.~(\ref{blkprop}) follows easily from an identity for the AdS bulk-bulk propagator $G_1^{(h)}(x,q)$, namely, that the action of the Laplacian on $x$ is equivalent to the action of the differential operator $Q_h$ on the temperature parameter $q$.

In this section, it will be useful to realize the bulk-bulk propagator in global AdS as the vacuum two-point function $ \langle 0|\Phi(x) \Phi(x')|0\rangle$ for a free quantum scalar field.\footnote{We use $\Phi(x)$ to indicate a bulk scalar field operator dual to the scalar quasiprimary $O(w)$ in the boundary CFT.}
Now let $(L_n,\overline{L}_n)$ be the isometry generators acting on the Hilbert space of the scalar field, as computed from Noether's theorem.   The operator implementing a translation around the thermal circle is $e^{2\pi i\tau L_0 -2\pi i \overline{\tau} \overline{L}_0} = q^{L_0}  \overline{q}^{\overline{L}_0}$ with $q=e^{2\pi i \tau}$.  Therefore, using the \sltwo invariance of the vacuum, the expression for the propagator whose endpoints are displaced by a single translation around the thermal circle is
\begin{equation}
G_1(x,q) =  \langle 0|\Phi(x)q^{L_0} \overline{q}^{\overline{L}_0} \Phi(x)|0\rangle~. \end{equation}
We now derive a differential equation for this object.  This analysis will only involve $q$ and not $\overline{q}$, so to avoid clutter we suppress the $ \overline{q}^{\overline{L}_0}$ insertion in what follows.

Using Eq.~(\ref{eq:bulkCasimir}-\ref{eq:bulkCasimir2}), we  have
\begin{equation}\label{nabG}
-{1\over 4}\nabla^2 G_1(x,q) = \langle 0| \Lc^2 \Phi(x) q^{L_0} \Phi(x)|0\rangle+ \langle 0| \Phi(x) q^{L_0}\Lc^2  \Phi(x)|0\rangle +2\eta^{AB}\langle 0|\Lc_A  \Phi(x) q^{L_0} \Lc_B \Phi(x)|0\rangle \,. \\
\end{equation}
Note that  $[L_n,\Phi]=-\Lc_n \Phi$, with the usual minus included so that the $\Lc_n$ obey the same algebra as the $L_n$. The first two terms are simple, but the cross-term requires some work to bring it to a more usable form.
We use $L_m q^{L_0} = q^{L_0+m}L_m$ to rewrite this last term as
\bea
2\eta^{AB}\langle 0|\Lc_A  \Phi(x) q^{L_0} \Lc_B \Phi(x)|0\rangle
&  = &-2\eta^{AB}\langle 0| \Phi(x)  q^{L_0+A}\Lc_B \Lc_A\Phi(x)|0\rangle\,.
\eea
Explicitly, there are three contributions
\begin{align}   -2\eta^{1,-1}\langle 0| \Phi(x)  q^{L_0+1}\Lc_{-1} \Lc_1\Phi(x)|0\rangle
&=  -q\langle 0| \Phi(x)  q^{L_0}(\Lc^2+L_0^2+L_0) \Phi(x)|0\rangle\,, \\
 -2\eta^{-1,1}\langle 0| \Phi(x)  q^{L_0-1}\Lc_{1} \Lc_{-1}\Phi(x)|0\rangle
&= -q^{-1}\langle 0| \Phi(x)  q^{L_0}(\Lc^2+L_0^2-L_0) \Phi(x)|0\rangle \,,\\
   -2\eta^{0,0}\langle 0| \Phi(x)  q^{L_0}\Lc_{0}^2 \Phi(x)|0\rangle &= 2\langle 0| \Phi(x)  q^{L_0}L_0^2 \Phi(x)|0\rangle \,.
\end{align}
which combine to give
\begin{align}
	2\eta^{AB}&\langle 0|\Lc_A  \Phi(x) q^{L_0} \Lc_B \Phi(x)|0\rangle \nonumber \\
& =-(q+q^{-1}) \langle 0| \Phi(x)  q^{L_0}\Lc^2 \Phi(x)|0\rangle -(q+q^{-1}-2)\langle 0| \Phi(x)  q^{L_0}L_0^2 \Phi(x)|0\rangle
\nonumber\\
&\qquad-(q-q^{-1}) \langle 0| \Phi(x)  q^{L_0}L_0 \Phi(x)|0\rangle \,,
\end{align}
a form that is useful in \cref{nabG}. Replacing insertions of $L_0$ by $q\p_q$ as before, and using the bulk free equation of motion $\Lc^2 \Phi = -h(h-1)\Phi$,
we finally arrive at
\begin{equation} -{1\over 4}\nabla^2 G_1(x,q) =(q+q^{-1}-2)h(h-1)G_1(x,q)-(q+q^{-1}-2)(q\p_q)^2 G_1(x,q) -(q-q^{-1})q\p_q G_1(x,q)    \end{equation}
or, more concisely,
\begin{equation}\label{Qeq} Q_h G_1(x,q) ={1\over 4}\nabla^2 G_1(x,q)  \end{equation}
where $ Q_h$ is the same differential operator appearing in the Casimir equation \cref{Qdef}.

\subsection{Application to torus 1-point block}
\label{holoonepoint}
Our proposed bulk representation of the torus one-point block is
\begin{equation}\label{eq:holoOnePt} W_1(h_\alpha,h_1;q) =  \mkern-24mu \int\displaylimits_\text{Thermal AdS}\mkern-24mu d^3x \sqrt{g} G_{b\p}^{(h_1)}(x,w)G_{1}^{(h_\alpha)}(x,q)~. \end{equation}
 We take the internal operator to be of dimension $h_\alpha$ and the external one to be of dimension $h_1$.  We now act with $Q_h(q)$ and use Eq.~(\ref{Qeq}).  Upon integrating by parts we have
\begin{equation}  Q_{h_\alpha} W_1(h_\alpha,h_1;q) = \mkern-24mu \int\displaylimits_\text{Thermal AdS}\mkern-24mu d^3x \sqrt{g} G_{1}^{(h_\alpha)}(x,q) {1\over 4} \nabla^2 G_{b\p}^{(h_1)}(x,w) = h_1(h_1-1)  W_1(h_\alpha,h_1;q)~, \end{equation}
which  matches the CFT equation Eq.~(\ref{onepointeq}).   It is also easy to see that our bulk expression has the small $q$ asymptotics $W_1(h_\alpha,h_1;q) \sim q^{h_\alpha}$ from the long-distance fall-off of the bulk to bulk propagator $G_1$.   This implies that $ W_1(h_\alpha,h_1;q) =  \mathcal{F}(h_{\alpha},h_1;q)$ up to an overall proportionality factor. The same derivation also applies to the representation in \cref{blkprop2}.
We have thus established our conjecture for the bulk representation of the torus 1-point block.

%%%%%%%%%%%%%%%%%
\section{Torus $n$-point function blocks}
\label{eq:torusncasimir}
In this section, we generalize the considerations of \cref{sec:torusonepointCasimir} to derive a Casimir equation satisfied by torus $n$-point-function blocks. We will show that in a particular channel, the block factorizes as a product of the one-point torus block and an $(n+2)$-point block on the sphere. Finally, we discuss the problem of giving holographic representations of these higher-point blocks.

\subsection{Casimir equations for $n$-point blocks}

As for the one-point functions, the $n$-point function on the torus can be decomposed into quasi-primary families labelled by $\alpha$:
\begin{equation}
 \Tr \left[ q^{L_0}\phi_1(w)\cdots \phi_n(w_n)\right]
=\sum_{\alpha}
\Tr \left[ P_{\alpha}q^{L_0}\phi_1(w_1)\cdots \phi_n(w_n)\right]\,.
\end{equation}
Unlike the $n=1$ case, the functional form of the terms in the decomposition is not yet determined kinematically. By taking the OPE between operators and decomposing into representations, or by inserting additional projections elsewhere in the trace, we may ultimately reduce the correlation function to sums of blocks determined by conformal symmetry in terms of only the conformal dimensions, with coefficients depending on quasiprimary OPE coefficients in the familiar way. There are several `channels', or ways to perform this procedure, but for now we will be ambivalent about the choice we have made, and define an $n$-point torus block
\begin{equation}
\mathcal{F}
(h_{\alpha};h_1,h_2,\ldots h_n;w_1,\ldots w_n;q)
=
\Tr \left[ P_{\alpha}q^{L_0}\phi_1(w_1)\cdots \phi_n(w_n)\right]\,
\end{equation}
where we have kept only the holomorphic dependence explicit, suppressed additional projectors onto conformal families, and ignored the coefficient that depends on dynamical data. None of this will effect the derivation of the Casimir equation which follows.

We may follow the same method as \cref{sec:torusonepointCasimir}, inserting a Casimir operator and commuting through the trace,  to arrive at the differential equation\footnote{To recover the torus one-point block Casimir equation of \cref{onepointeq} from this equation, use the fact that $\mathcal{L}_0$ annihilates $\mathcal{F}$, and then because there are no cross-terms in the sum, the second term is just the Casimir differential operator: $\mathcal{L}_{+1}^{(1)} \mathcal{L}_{-1}^{(1)}\mathcal{F} = (\mathcal{L}^{(1)})^2 \mathcal{F} = -h_1(h_1-1) \mathcal{F}$.}
\begin{equation}
\left[Q_{h_\alpha}
+\sum_{i=1}^n \calL_{+1}^{(i)} \sum_{j=1}^n \calL_{-1}^{(j)} \right] \mathcal{F}
=0 \,,
\end{equation}
where the $\mathcal{L}_n^{(i)}$ are differential operators \cref{eq:cylLn} acting on $w_i$, and $Q_{h_\alpha}$ is the differential operator acting on $q$, exactly as in \cref{Qdef}:
\begin{equation}
Q_{h} =
q(1-q)^2 \partial_q^2
-2(1-q)q \partial_q
-h(h-1)\frac{(1-q)^2}{q}~.
\end{equation}
Now, we define `total' differential operators
\begin{equation}
\calL^{tot}_{A}\equiv \sum_{i=1}^n \calL_{A}^{(i)}\quad , \quad A=0,\pm1,
\end{equation}
acting on all the $e_i$, equivalent to the insertion of the operator $L_n$ on a cycle surrounding all the $\phi_i$, such that the differential equation is succinctly written as
\begin{equation}\label{eq:npointblockDiffEq0}
\left[
Q_{h_\alpha}
+\calL^{tot}_{+1}\calL^{tot}_{-1}
\right]{ \cal F}
=0\,.
\end{equation} Furthermore, by inserting $L_0$ into the trace and commuting $L_0$ through together with using the cyclicity property of the trace we have translation invariance $\calL_0^{tot} \mathcal{F} =0$. This implies that
\begin{equation}
\left(\calL^{tot}\right)^2 = \eta^{AB} \calL_A^{tot} \calL_B^{tot} \mathcal{F}
=
\left[
-\calL^{tot}_0 (\calL^{tot}_0+1)
+\calL_{+1}^{tot} \calL_{-1}^{tot}
\right]
\mathcal{F}
= \calL^{tot}_{+1} \calL^{tot}_{-1} \mathcal{F}.
\end{equation}  With this, we can rewrite the differential equation as
\begin{equation}\label{eq:npointblockDiffEq}
\left[
Q_{h_\alpha}
+\left(\calL^{tot}\right)^2
\right]\mathcal{F}
=0\,.
\end{equation}
The second term $\left(\calL^{tot}\right)^2$ is the same differential Casimir operator that appears when deriving the sphere $(n+2)$-point conformal block\cite{Dolan:2000ut}, appearing in the sphere correlator $\langle \phi_\alpha \phi_1\ldots \phi_n \phi_\alpha\rangle$.

\begin{figure}[h!]
 \centering
  \includegraphics[width=0.5\textwidth]{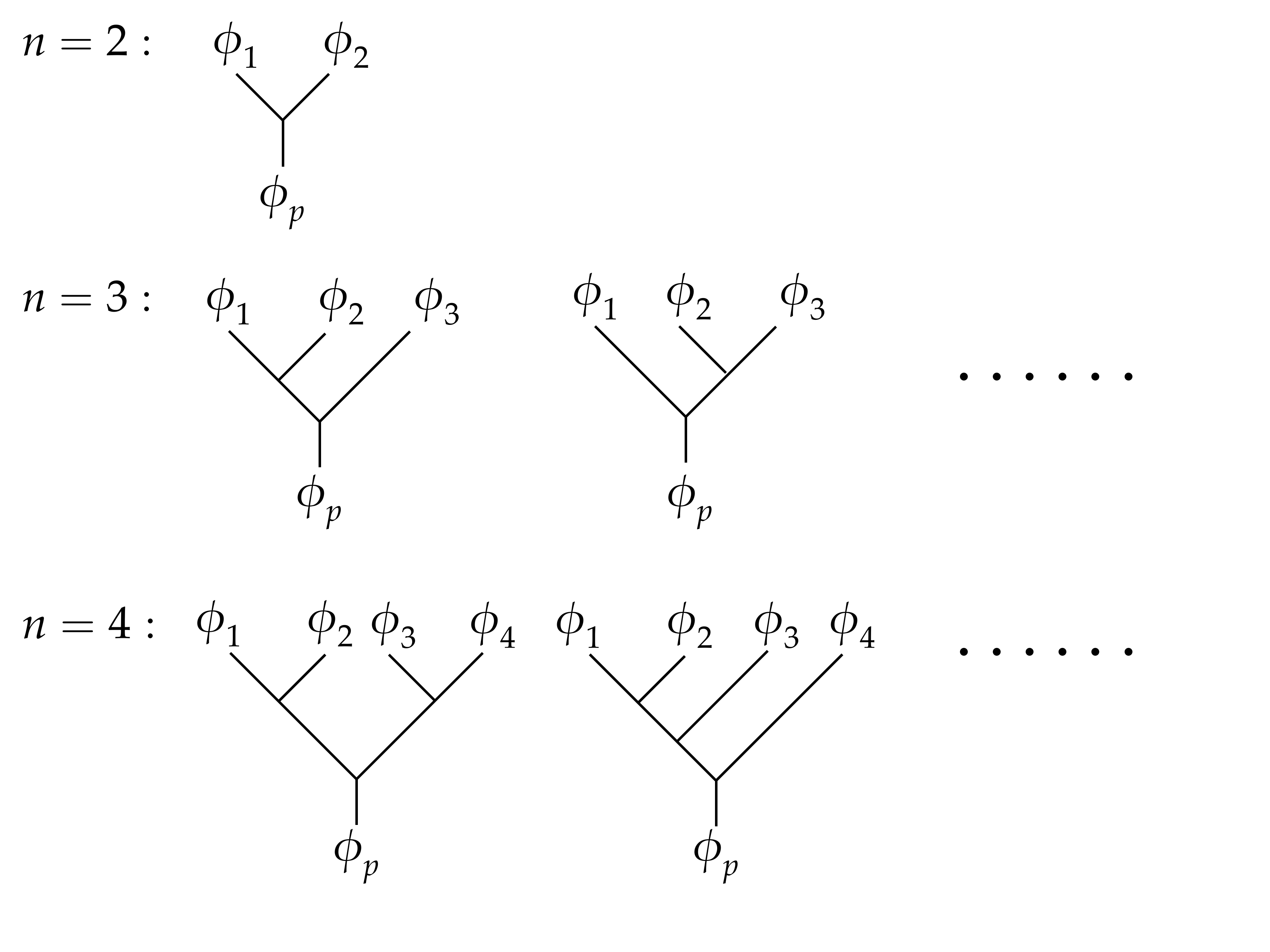}
 \caption{
Given a particular OPE channel, $\phi_p$ is the last operator appearing in the OPE.
 \label{fig:Op}
}
 \end{figure}

Now, suppose we have chosen a channel where we do not insert any other projection operators in the trace, but rather only take the OPE repeatedly as for correlation functions on the plane. At the last stage, once we have taken the OPE with every pair of operators, the block contains contributions only from the conformal family of some primary $\phi_p$  (see \cref{fig:Op}). In other words, the block includes a projection operator $P_p$ on the cycle surrounding all the $w_i$. This means that the differential operator $\left(\calL^{tot}\right)^2$ acting on the coordinates $w_i$ is just the Casimir of that representation, so it can be replaced by the constant $-h_p(h_p-1)$. The block satisfies the same differential equation in $q$ alone as the one-point block, and fixing the solution using the low-temperature asymptotics, this implies that the dependence of $q$ and $w_i$ factorizes:
\begin{equation}
\mathcal{F}(h_{\alpha};h_1,h_2,\ldots h_n;w_1,\ldots w_n;q) = \mathcal{F}(h_{\alpha},h_p;q)\; \mathcal{F}_{n+2}(h_p,h_p,h_i;w_i)~.
\end{equation}
Here, the first factor is the one-point torus block from above, and the second factor is just the $(n+2)$-point block on the cylinder at zero temperature, with two insertions of the operator $\phi_p$, at $t=\pm\infty$.

For example, the two-point block in this channel factorizes as the one-point torus block times the more standard four-point block, with a well-known expression in terms of a hypergeometric function:
\begin{align}\label{eq:twopointblock1}
	\mathcal{F}(h_\alpha;h_1,h_2;w_1,w_2;q)&= \frac{q^{h_\alpha}}{(1-q)^{1-h_p}} \: {}_2F_1(h_p,h_p+2h_\alpha-1;2h_\alpha;q)  \nonumber\\
	&  \qquad \times (1-z)^{-h_1-h_2+h_p} \: _2F_1\left(h_p,h_p-h_1+h_2;2 h_p;1-z\right) \,.
\end{align}
The first line is the torus one-point block from the previous section, while the second line is the (holomorphic) T-channel block for $\langle \phi_\alpha \phi_1\phi_2 \phi_\alpha\rangle$, with cross-ratio $z\equiv e^{-i (w_2-w_1)}$.

In more general channels, with additional projection operators inserted in the trace, there will not be such a simple factorization, and the solution to the Casimir equation must be written as a sum over many such factorized pieces with different eigenvalues. It is possible to get additional Casimir equations in such cases, but we will leave considerations of these other channels for future work.

\subsection{Holographic description of a torus 2-point block}

The holographic representations of higher-point blocks are more subtle than the one-point block. This is because the Witten diagram contains contributions from double-traces built from the external operators, coming from contact terms in bulk propagators, which are hard to project out in a natural way. This is in contrast to the contributions from multi-trace operators built from the internal conformal family, which are simply and naturally projected out by replacing the full thermal AdS propagator with $G_1$. In this section, we discuss possible proposals and the associated difficulties in detail, ending with a tentative suggestion for a representation of the two-point thermal block.

\subsubsection*{Building a bulk representation}

Given the intuition from previously known results for geodesic Witten diagrams, and the representation of the one-point torus block, a natural ansatz to write down for a bulk representation of a two-point torus block is the following:
\begin{equation}
\label{eq:HoloTwoPt}
W_2(q;w_1,w_2) =
\int_{AdS} \! d^3x \sqrt{g}  ~
G^{(h_\alpha)}_1(x,q)
\int_{ \gamma_{12}} d \lambda \;
\tilde{G}^{(h_p)}_{bb}(x;y(\lambda))
G^{(h_1)}_{b\p}(w_1;y(\lambda))
G^{(h_2)}_{b\p}(w_2;y(\lambda))
 \end{equation}
Here, $G_1$ is the bulk propagator used for the one-point block, $\tilde{G}^{(h_p)}_{bb}$ is some `bulk-to-bulk propagator', the exact form of which we will discuss, and $y$ is a bulk point, which we have chosen to integrate over the geodesic $\gamma_{12}$ between points $w_1$ and $w_2$, following the example of geodesic Witten diagrams and the `OPE block' introduced in \cite{Czech:2016xec}, and discussed in more detail later. For the consideration of the Casimir equation which follows, nothing would change if we were to integrate $y$ over the whole bulk.

Given this expression $W_2$, let us attempt to derive the Casimir equations for the two point block. First we act with the differential operator $Q_{h_\alpha}$ on the $q$ variable, and similarly to the derivation in the holographic one-point block in \cref{holoonepoint}, we use the bulk propagator identity \cref{Qeq} to convert this to a Laplacian acting on $x$. After integrating by parts, the Laplacian acts on $\tilde{G}^{(h_p)}_{bb}$, which results in the Casimir equation if $\tilde{G}^{(h_p)}_{bb}(x;y)$ obeys the free wave equation as a function of $x$, \emph{without sources}:
\begin{align}
	Q_{h_\alpha}W_2(q;w_1,w_2) &=
\int_\text{AdS} \! d^3x \sqrt{g}  \:
\left[Q_{h_\alpha}  G^{(h_\alpha)}_1(x,q;h_\alpha) \right]
\int d\lambda\: \tilde{G}^{(h_p)}_{bb} G^{(h_1)}_{b\p} G^{(h_2)}_{b\p}  \nonumber\\
&=
\int_{AdS} \! d^3x \sqrt{g}  ~
G^{(h_\alpha)}_1
\int d\lambda\;
\left[\left(\frac{1}{4}\nabla_x^2\right) \tilde{G}^{(h_p)}_{bb}(x;y)\right]
G^{(h_1)}_{b\p}
G^{(h_2)}_{b\p}
\nonumber\\
&=h_p(h_p-1) W_2(q;w_1,w_2)
\end{align}
	If $\tilde{G}^{(h_p)}_{bb}$ did not obey the wave equation, but instead had some source in the bulk (for example, the usual bulk-to-bulk propagator would have a delta-function source at $y$), we would not recover the Casimir equation. This would give contact terms in the bulk integral, which provide the contribution of double-trace operator exchanges in the full Witten diagram.

Next, similar to the Casimir-equation derivation of the holographic representation of the sphere four-point block \cite{Hijano:2015zsa}, we can rewrite
\begin{equation}
W_2 = \int \! d^3x \sqrt{g}  \:
G^{(h_\alpha)}_1(x,q;h_\alpha)
F(x;w_1,w_2)
\end{equation}
where $F$ is defined as\footnote{Note that $F$ here is different from the $F$ in \cite{Hijano:2015zsa} since we use the modified source-free bulk-to-bulk propagator $\tilde{G}_{bb}^{(h_\alpha)}$ instead of an ordinary AdS bulk-to-bulk propagator $G_{bb}^{(h_\alpha)}$.}
\begin{equation}
F(x;w_1,w_2)\equiv \int_{\gamma_{12}} d \lambda\: \tilde{G}^{(h_p)}_{bb}(x,y(\lambda))  G^{(h_1)}_{b\p}(w_1;y(\lambda)) G^{(h_2)}_{b\p}(w_2;y(\lambda)).
\end{equation}
If we assume that the combination $F$ is invariant under AdS isometries acting simultaneously on $w_1$, $w_2$ and $x$, then it satisfies
\begin{equation}
\calL_A^{(tot)}F(x;w_1,w_2)
=\left[
\calL_A^{(1)}
+\calL_A^{(2)}
\right] F(x;w_1,w_2)
=- \calL_A^{(x)} F(x;w_1,w_2)
\end{equation}
where the operators $L_A^{(i)}$ act on the coordinates $w_i$ while $L_A^{(x)}$ acts on the bulk coordinate $x$. Acting with $L_A^{(tot)}$ again and summing over $A=0,\pm 1$ yields
\begin{equation*}
\sum_A \calL_A^{tot}\calL_A^{tot} F(x;w_1,w_2)
=\sum_A\left(\calL_A^{(x)}\right)^2 F(x;w_1,w_2)
=-\frac{1}{4} \nabla_x^2 F(x;w_1,w_2) =-h_p(h_p-1) F(x;w_1,w_2)
\end{equation*}
where we once again use the free equation of motion for $\tilde{G}^{(h_p)}_{bb}$. Since all $w_i$ dependence of $W_2$ is contained in $F$, this establishes the second Casimir equation for the block.

These conditions on $\tilde{G}^{(h_p)}_{bb}$ are not sufficient to show that the expression $W_2$ really is the two-point torus block, since the Casimir differential equations do not have unique solutions without also providing boundary conditions. For example, we might choose $\tilde{G}^{(h_p)}_{bb}$ to be the usual bulk-to-bulk propagator, minus its `shadow', the Green's function with alternate boundary conditions as relevant for a dimension $1-h_p$ operator. The bulk sources in the two terms cancel, so the Casimir equation would be satisfied, but the result would not be a single block, but a linear combination including the shadow block.

In summary, to find a bulk representation of a conformal block, we see two possible obstacles. Firstly, contact terms in bulk to bulk propagators give unwanted double-trace contributions, and secondly, the wrong boundary conditions give shadow block contributions, and both of these must be avoided. For example, the geodesic Witten diagram for four-point blocks uses the usual bulk-to-bulk propagator to avoid the second problem, and avoids contact terms by integrating only over geodesics. It is not obvious how to generalize this to higher point blocks.

\subsubsection*{A proposal from the OPE block}

One way to think of the two-point thermal block in \cref{eq:twopointblock1} is as a trace in the representation built on conformal dimension $h_\alpha$, of the `OPE block' $[O_1(w_1)O_2(w_2)]_p$ discussed in \cite{Czech:2016xec}, which packages the conformal family of $O_p$ appearing in the $O_1,O_2$ OPE. This can be written in Lorentzian signature as a smearing of $O_p$ over the causal diamond bounded by the spacelike separated points $w_1,w_2$, with an appropriately chosen kernel. This has a natural bulk description as a free bulk field $\Phi_p(x)$ integrated over a geodesic, where $\Phi_p$ is defined using the `HKLL' reconstruction \cite{Hamilton:2006az,Hamilton:2006fh} of the free bulk field. This writes $\Phi_p(x)$ as an integral of the operator $O_p$ on the boundary by using the smearing function $K^{(h_p)}_{HKLL}$, supported in the causal diamond: $\Phi_p(x) =\int_\diamond\! d^2w K^{(h_p)}_{HKLL}(x,w) O_p(w)$. This gives us the following representation for the OPE block:
\begin{equation}
[O_1(w_1)O_2(w_2)]_p  = \int_{\gamma_{12}} \! d\lambda G^{(h_1)}_{b\p}(w_1,y(\lambda)) G^{(h_2)}_{b\p}(h_2,w_2,y(\lambda))\int_\diamond\! d^2w K^{(h_p)}_{HKLL}(y(\lambda),w) O_p(w)~.
\end{equation}
Note that this is just a CFT operator equation, valid in correlation functions with other operators inserted outside the causal diamond, though expressed in bulk language.

If we take the matrix elements of this expression between some quasiprimary states $|\alpha\rangle,|\beta\rangle$, the right hand side gives $\langle\beta|O_p(w)|\alpha\rangle$, the dependence on $w$ being a simple kinematically determined function depending only on $h_\alpha-h_\beta$, integrated against the smearing kernel. Performing the integrals then results in the four-point block with intermediate operator $O_p$, which can be checked by doing the integrals with an explicit expression for the OPE block.

Given this result, and the factorized form for the two-point torus block, a natural suggestion is to use the HKLL kernel to build the modified bulk-bulk propagator
\begin{equation}
\tilde{G}^{(h_p)}_{bb}(x,x')
=\int_{\partial AdS_3} d^2 w ~
G^{(h_p)}_{b\partial}(x,w)
K^{(h_p)}_{HKLL}(x',w),
 \end{equation}
by multiplying the bulk-boundary propagator against the smearing function and integrating over the common boundary point.  This obeys the source free wave equation $(\nabla^2 -m_p^2)\tilde{G}=0$ in $x$, and as a boundary condition, the coefficient of the non-normalizable mode approaches the HKLL function. Note that $\tilde{G}^{(h_p)}_{bb}$ here depends implicitly on $w_1,w_2$ through the choice of integration region and HKLL kernel, and is only defined when $x'$ lies on the geodesic $\gamma_{12}$. The expression is rather formal as it stands, because in the range of integration $w$ must be Lorentzian, but the finite-temperature interpretation requires $x$ to be allowed to be a point in Euclidean AdS.

Given this form of $\tilde{G}^{(h_p)}_{bb}$, the factorized form \cref{eq:twopointblock1} of the block follows directly from the integral representation. First, we note that the $q$ dependence of $W_2$ is now identical to the expression for the one-point block \cref{blkprop2}, since the $x$ dependence of $\tilde{G}^{(h_p)}_{bb}$ now comes through the bulk-to-boundary propagator (albeit to a formally Lorentzian boundary point). The boundary point $w$ is later integrated over the causal diamond, but this apparent $w$ dependence is irrelevant, since the one-point block is in any case $w$ independent by time translation and rotation symmetry. The block therefore factorizes, with the remaining $w$-dependent factor being
\begin{equation}
\int_{ \gamma_{12}} d \lambda
G^{(h_1)}_{b\p}(w_1;y(\lambda))
G^{(h_2)}_{b\p}(w_2;y(\lambda))\,
\int_{\partial AdS_3} d^2 w \:
K^{(h_p)}_{HKLL}(w;y(\lambda))\; .
\end{equation}
This is nothing other than the expression for the four-point block coming from the expectation value of the OPE block in some state, as discussed above but with $|\alpha\rangle=|\beta\rangle$, in which case $\langle\alpha|O_p(w)|\alpha\rangle$ is a constant, independent of $w$. The choice of state does not come into this expression, but this is not unexpected since the four-point block itself is independent of $h_\alpha$ (depending only on the difference in dimension between the two states appearing, which is zero here).

At this stage, our bulk representation of the two-point torus block is somewhat formal, as it involves mixed Euclidean and Lorentzian signatures. In particular, it would appear to be problematic when the bulk point $x$ crosses the lightcone of $w$, because of lightcone singularities in the bulk to boundary propagator.  Similarly, the convergence properties of the integrals are unclear. It would be interesting to understand these issues better in order to make this proposal more rigorous.

%%%%%%%%%%%%%

\section{Bulk geodesic description of torus blocks in the heavy limit}

Conformal blocks simplify in the limit in which all operator dimensions, internal and external, become large, $h \gg 1$.  In particular, the conformal blocks exponentiate in this regime.  Based on existing results (see \cite{Hijano:2015zsa}), we expect that the function appearing in the exponent is equal to the action of a network of bulk worldlines in AdS, whose configuration is taken to minimize the total worldline action, as obtained from an eikonal approximation. In this section we will verify this correspondence in full generality.
We begin with the sphere $n$-point block, showing how the action for the network of worldlines obeys the same conformal Casimir equation as does the conformal block.   We then consider the extension to the torus $n$-point block.

Before proceeding, let us note that global conformal blocks in the limit $h \gg 1$ are expected to coincide with Virasoro blocks in the semiclassical limit in which $ c \rightarrow \infty$, $h\rightarrow \infty$, with $h/c$ held fixed but considered to be small, $h/c \ll 1$ \cite{Fitzpatrick:2015zha}.  This correspondence is easily understood from the bulk AdS point of view, where in both cases there is a natural correspondence with non-gravitating particle worldlines.

\subsection{Sphere block}

\subsubsection{Large dimension limit in CFT}

\begin{figure}
  \centering
  \includegraphics[width=5cm]{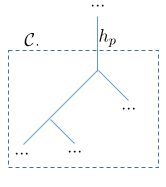}\\
  \caption{
  An OPE channel where all the operators appearing to the right of $P_p$ (i.e. $\phi_1,\ldots, \phi_j$) are grouped inside the blue square.
  }\label{blocksphere}
\end{figure}
We begin by studying the decomposition of an $n$-point function on the complex plane  in a particular OPE channel:
\begin{align}
\langle \phi_n(z_n)\phi(z_{n-1}) \ldots \phi_1(z_1)\rangle
&=\sum_{p}
\langle \phi_n(z_n)\phi(z_{n-1}) \ldots \phi_{j+1}(z_{j+1}) ~P_p ~ \phi_j(z_{j})\ldots \phi_2(z_{2})\phi_1(z_1)\rangle
\cr &\equiv \sum_p |\mathcal{F}_p|^2
\end{align}
where this may be thought of in radial quantization, with all $z_i$ for $i\leq j$ inside a circle on which we insert the projector $P_p$, and those for $i>j$ outside the circle. There may be other projections inserted, which we suppress. We consider the large conformal dimension limit $h_i ,h_p\rightarrow \infty$, with ratios $h_i/h_p$ fixed. The subscript $p$ labels the global conformal family while  $j$ tells us how many operators sit to the right of $P_p$.
Furthermore, we perform an OPE expansion on all operators to the right of $P_p$, keeping a single representation at each step,  such that the operator $\phi_p$ appears in the last OPE step
(see Fig.~\ref{blocksphere}).

Next, it is useful to introduce Ward identities \cite{BPZ} for the insertions of conformal generators
\begin{equation}
 \mathcal{F}_p(L_m^{\cal{C}})\equiv\langle \ldots  L_{m}P_p \ldots\rangle
=\sum_{i=1}^{j}\left(
(1+m)h_{i}z^{m}_{i}
+z_{i}^{m+1}\frac{\partial}{\partial z_{i}}
\right)\langle \ldots P_p \ldots\rangle,
\end{equation}
or more explicitly,
\begin{align}
	\label{eq:semiWard}
 \frac{\mathcal{F}_p(L_{-1}^{\cal{C}})}{\mathcal{F}_p} &= \sum_{i=1}^{j}\frac{\partial}{\partial z_i}\log \mathcal{F}_p \notag \\
\frac{\mathcal{F}_p(L_0^{\cal{C}})}{\mathcal{F}_p} &= \sum_{i=1}^{j}\left(h_{i}+z_{i}\frac{\partial}{\partial z_i}\log \mathcal{F}_p\right) \notag \\
\frac{\mathcal{F}_p(L_1^{\cal{C}})}{\mathcal{F}_p} &= \sum_{i=1}^{j}\left(2h_{i}z_{i}+z_{i}^2\frac{\partial}{\partial z_{i}}
\log\mathcal{F}_p\right),
\end{align}
where $\mathcal{F}_p = \mathcal{F}_p(1)$ is just the block itself.
The notation ${\cal C}$ specifies that in radial quantization the \sltwo conformal generators $L_A^{\cal C}$ act on operators inside the contour of integration ${\cal{C}}$ (i.e. operators $\phi_1,\ldots, \phi_j$) defining the moments of the stress tensor. This is depicted by the dashed square in Fig.~\ref{blocksphere}.

We also know that inserting the Casimir operator along with the projection gives
\begin{equation}
\frac{\langle \ldots \left[\frac{1}{2}(L_1L_{-1}+L_{-1}L_1)-L_0^2 \right]P_p \ldots \rangle}
{\langle \ldots P_p \ldots\rangle}=-h_{p}(h_{p}-1). \end{equation}

We now make the ansatz that the block exponentiates in the limit of large dimensions,
\begin{equation} \mathcal{F}_p \approx e^{-S}, \end{equation}
where $S$ scales linearly in the dimensions, of order $h_p$ or $h_i$. Keeping only the leading order in the limit, the Casimir equation then simplifies as
\begin{equation}
\label{eq:semiCasimir} \frac{\mathcal{F}_p(L_1^{\cal{C}})}{\mathcal{F}_p}\frac{\mathcal{F}(L_{-1}^{\cal{C}})}{\mathcal{F}_p}
-\left(\frac{\mathcal{F}_p(L_0^{\cal{C}})}{\mathcal{F}_p}\right)^2=-h_p^2
\end{equation}
because we keep only terms where derivatives act on the block itself, and bring down a factor of dimension, rather than acting on factors from the action of previous $L_n$s.

Next, we will show that the Ward identities \cref{eq:semiWard} and hence the semiclassical Casimir equation \cref{eq:semiCasimir} are obeyed by the action of a network of particle geodesics in AdS.

\subsubsection{Geodesic networks in AdS}
\label{sec:adsnetwork}
In this section we consider AdS in Poincare coordinates,
\begin{equation}
ds^2 = {du^2 + dz d\zb \over u^2}~
\end{equation}
with Killing vectors
\begin{align}
	\Lc_{-1} &= -\p_z \cr
\Lc_0 &= - (z\p_z +{1\over 2} u\p_u)  \cr
\Lc_1 &= - ( z^2 \p_z +z u\p_u -u^2 \p_{\zb} )
\end{align}
obeying the \sltwo algebra $[\Lc_{m},\Lc_{n}]=(m-n)\Lc_{m+n}$.

We now consider a network of geodesic segments in AdS.  A worldline is taken to emanate from the location of each external operator location on the boundary, and we then connect them in the bulk using cubic vertices to form a network that mimics the particular OPE channel considered in the definition of the analogous conformal block.   That is, we push the OPE diagram \cref{blocksphere} into the bulk, holding fixed the locations of the external operators on the boundary.  The action of the network is given by summing the over the lengths of the segments weighted by twice the conformal dimension $2h$ (which is equal to  the bulk mass in the heavy limit) of the corresponding operator.  The on-shell action is given by extremizing with respect to the worldline trajectories (which are geodesics) and the locations of the vertices.  We then wish to show that this on-shell action computes the conformal block in the heavy limit via
\begin{equation}\label{onshell1}
\log\mathcal{F}_p=-S_\text{on-shell}\left(\{z_i,\bar{z}_i,u_i^{(\infty)}\}\right),
\quad \text{where}\quad
S_{\text{on-shell}}\left(\{z_i,\bar{z}_i,u_i^{(\infty)}\}\right)=\sum_{{\rm segments}} 2h_\alpha l_\alpha,
\end{equation}
where the $l_\alpha$ are the appropriate geodesic lengths,  including both the bulk-to-bulk and bulk-to-boundary geodesics.  The geodesics lengths diverge in going to the boundary at $u=0$, and so we have imposed a cutoff $u^{(\inf)}$.  We should properly deal with a renormalized action obtained by subtracting off the divergence, but this just contributes an overall $z_i$-independent factor, and so we will suppress this.

The total geodesic network does not have loops, so for any worldline we may split it into two parts, joined together by that geodesic. The two parts connect up to boundary points $z_{1}, \ldots z_j$ and $z_{j+1}, \ldots z_n$ respectively, and are joined by a worldline carrying dimension $h_p$.  This structure is chosen to coincide with the OPE channel chosen in the CFT.

We now compute the variation of the on-shell action
generated by the action of a Killing vector $K$ on the locations of the external operators at $x_1, \ldots x_j$ where $x_i\equiv (z_i,u_i^{(\infty)})$.\footnote{If we acted with the Killing vector on the locations of all boundary points $x_1, \ldots x_n$ the total action would be invariant by symmetry.}  Since $S_{\text{on-shell}}$ is a scalar function of the $x_i$'s, it transforms as
 \begin{equation}
\delta_K S_{\text{on-shell}}(\{x_i\})=
 \sum_{i=1}^{j} {\cal{K}}^{(i)} S_{\text{on-shell}}(\{x_i\}),
  \end{equation}
where ${\cal{K}}^{(i)}$ is the Killing vector acting on the $x_{i}$ coordinate as
\begin{equation} {\cal{K}}^{(i)}=K^\mu(x^i) \frac{\partial}{\partial x^{\mu}_{i}}. \end{equation}
On the other hand, because of the on-shell condition, the variation of the action is a boundary term,\footnote{\label{fn1}This follows from the standard derivation of the geodesic equation by extremizing  the worldline action $l$, but keeping track of the boundary term, which is given by $- g_{ab}\frac{dX^a}{ds}\delta X^b$ where $X^a(s)$ is the geodesic parametrized by $s$ and $\delta X^a$ is the variation. Defining the tangent vector $\hat{l}^a\equiv \frac{dX^a}{ds}$ and setting $\delta X^\mu =K^\mu$ give Eq.~(\ref{eq:Sonshell}).}
\begin{equation}\label{eq:Sonshell}
\delta_K S_{\text{on-shell}}(\{x_i\})=
\sum_{i=1}^j
2h_{i}
\langle {\cal K},\hat{l}_{i}
\rangle(x_i)
,
 \end{equation}
where $\hat{l}_{i}$ denotes the unit tangent vector of the geodesic with $x_{i}=(z_i,u_i^{(\infty)})$ as the end point.
Here and subsequently, $\langle \cdot , \cdot \rangle$ denotes the inner product in the AdS metric, i.e. $\langle  A ,B \rangle(x)
\equiv g_{\mu\nu}(x) A^\mu(x) B^\nu(x)$ for vectors $A^\mu(x)$ and $B^\nu(x)$.

We now note two facts.  First,  since $K$ is a Killing vector, the inner product $\langle{\cal{K}},\hat{l}_i\rangle$ is a constant along any geodesic segment.
Second, extremization of the action imposes a local condition on the tangent vectors at each cubic vertex, which can be thought of as `balancing the forces' between the three geodesics,
\begin{equation}
\left. \sum_{k=1}^{3}h_k{\hat{l}}_k \right|_v=0~,
 \end{equation}
where the tangent vectors $\hat{l}_k$ all point out from a bulk vertex $v$.
 These two properties imply that we can express Eq.~(\ref{eq:Sonshell}) in terms of data of the $h_p$ worldline,
 \begin{equation}
 \delta_K S_{\text{on-shell}}(\{x_i\})=-2h_p\langle{{\cal{K}}},\hat{l}_p\rangle(v_p)
 \end{equation}
where $\hat{l}_p$ is the unit vector pointing out of the vertex $v_p$ connected by the particle worldline with dimension $h_p$. Comparing the two variations of the action, we have
\begin{equation}
\sum_{i=1}^{j} {\cal{K}}^{(i)} S_{\text{on-shell}}(\{x_i\})=-2h_p\langle {\cal{K}},\hat{l}_p \rangle(v_p).
\end{equation}
Using the explicit form of the Killing vectors, we can evaluate this equation for the  \sltwo generators:
\bea\label{Killingdiff} &~& 2h_p\langle {\cal{L}}_{-1},\hat{l}_p\rangle=\sum_{i=1}^{j}
\left[\frac{\partial}{\partial z_{i}}S_{\text{on-shell}} \right]\notag \\
&~& 2h_p\langle{\cal{L}}_0,\hat{l}_p\rangle=\sum_{i=1}^{j}
\left[
z_{i}\frac{\partial}{\partial z_{i}}S_{\text{on-shell}}+\frac{1}{2}u_{i}^{(\inf)}
\frac{\partial}{\partial u_{i}^{(\inf)}}S_{\text{on-shell}} \right]\notag \\
&~&2 h_p \langle {\cal{L}}_1,\hat{l}_p\rangle=\sum_{i=1}^{j}\left[
z_{i}^2\frac{\partial}{\partial z_{i}}S_{\text{on-shell}}-\left(u_{i}^{(\inf)}\right)^2
\frac{\partial}{\partial \zb_{i}}S_{\text{on-shell}}+z_{i}u_{i}^{(\inf)}
\frac{\partial}{\partial u_{i}^{\inf}}S_{\text{on-shell}}\right] \,.\notag \\
\eea
Note that the LHS is evaluated at the bulk vertex $v_p$ while the RHS is evaluated at the boundary points $\{x_i\}$.

To evaluate the derivatives with respect to the cutoff, we note that the action for the geodesic segments approaching the boundary diverge logarithmically as $S \sim -2h_i \log u_i^{(\infty)}$, and so we have $\scriptsize{u_{i}^{(\infty)}}\frac{\partial}{\partial u_{i}^{(\infty)}}S_{\text{on-shell}}= -2h_i$. Substituting this into Eq.~\ref{Killingdiff} and sending the cutoffs $u_{i}^{(\infty)}$ to zero gives
\bea
 &~&-2h_p\langle {\cal{L}}_{-1},\hat{l}_p\rangle=\sum_{t=1}^{j}\frac{\partial}{\partial z_{i_t}}\left(-S_{\text{on-shell}}\right) \notag \\
&~&-2h_p\langle{\cal{L}}_0,\hat{l}_p\rangle=\sum_{i=1}^{j}\left[
h_{i}
+z_{i}\frac{\partial}{\partial z_{i}}\left(-S_{\text{on-shell}}\right)\right] \notag \\
&~&
-2h_p \langle {\cal{L}}_1,\hat{l}_p\rangle=\sum_{i=1}^{j}
\left[
2h_{i}z_{i}
+
z_{i}^2\frac{\partial}{\partial z_{i}}\left(-S_{\text{on-shell}}\right)
\right].
\eea These reproduce the Ward identities in Eq.~(\ref{eq:semiWard}) if we identify
\bea \label{eq:identification}
&~& \log\mathcal{F}=-S_\text{on-shell}\left(\{z_i,\bar{z}_i,u_i^{(\infty)}\}\right)+{\text{constant}} \notag \\
&~&
\frac{\mathcal{F}(L_A)}{\mathcal{F}}=-2h_p\langle {\cal{L}}_A,\hat{l}_p\rangle(v_p)\,.
\eea Note that interestingly in the RHS, the object in the first line depends on purely  boundary points (as it should) while the object in the second line is evaluated at the bulk vertex $v_p$.

Lastly, since at each point in AdS, the  vectors ${\cal{L}}_A$'s  form a complete basis of  vectors, satisfying
\begin{equation}
4\left(\langle {\cal{L}}_0,\hat{l}_p\rangle^2-\langle{\mathcal{L}}_1,\hat{l}_p\rangle\langle{\mathcal{L}}_{-1},\hat{l}_p\rangle\right)=\langle \hat{l}_p,\hat{l}_p\rangle=1\,, \end{equation}
under the identifications of \cref{eq:identification}, we recognize this as giving the Casimir in \cref{eq:semiCasimir}.

In summary, we have shown that geodesic networks obey the same semiclassical Casimir equations as blocks at large dimension in CFT.
As a pair of external operators are brought together, the geodesic network also share the same behavior as boundary conditions for the Casimir equations in the CFT. This establishes that the two quantities are equal, up to a unimportant overall factor.

\subsection{Torus blocks}

Building on the previous subsection, we now present the holographic description of global blocks on the torus in the large conformal dimension limit.

\subsubsection{Field theory}

 \begin{figure}
 \centering
	\begin{subfigure}[b]{.4\textwidth}\centering
	 \includegraphics[width = 0.7\textwidth]{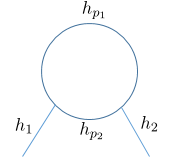}
	 \caption{Projection channel\label{ProjChan}}
	\end{subfigure}
	\begin{subfigure}[b]{.4\textwidth}\centering
	 \includegraphics[width = 0.55\textwidth]{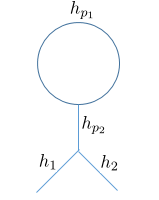}
	\caption{OPE channel\label{OPEChan}}
	\end{subfigure}
\caption{\label{Torus2pt}Two different channels for two point function conformal blocks on the torus, with either a projection inserted between the operators, or the OPE taken between them.}
 \end{figure}

As discussed in Sec.~\ref{eq:torusncasimir}, there are various channels channels depending on where the projection operators are inserted.
As an example, for torus two-point blocks, there are two possible channels, as illustrated in \cref{Torus2pt}.
We shall focus on torus multi-point blocks  such as that  in \cref{OPEChan}, where we first perform the OPE of all external operators.\footnote{See the discussions below Eq.~(\ref{eq:npointblockDiffEq}) for a more detailed description of this channel.}

Recall that the global torus $n$-point block can be defined as
\begin{equation}
\mathcal{F}_p=\Tr\left[ P_{p}q^{L_0}\phi_1(w_1)\phi_2(w_2) \ldots \phi_n(w_n) \right]
\end{equation}
where we have left implicit the OPE decomposition of the string of operators.
Let us now define
\bea
&~& \mathcal{F}^{(u)}(L_A) \equiv
\Tr\left[ P_{p}  q^{L_0}L_A\phi_1(w_1)\phi_2(w_2) \ldots \phi_n(w_n) \right]
\notag \\
&~& \mathcal{F}^{(d)}(L_A)
\equiv
\Tr\left[ P_{p}  L_Aq^{L_0} \phi_1(w_1)\phi_2(w_2) \ldots \phi_n(w_n) \right]
 . \eea
Using Eq.~(\ref{eq:useful}), we have
\bea &~&\mathcal{F}^{(u)}(L_0)=\mathcal{F}^{(d)}(L_0)=q\frac{\partial}{\partial q}\mathcal{F}_p \notag \\
&~&\mathcal{F}^{(u)}(L_{-1})=q\mathcal{F}^{(d)}(L_{-1}) \notag \\
&~&\mathcal{F}^{(u)}(L_1)=\frac{1}{q}\mathcal{F}^{(d)}(L_1).
\eea
On the other hand, from the commutation relations and the cyclicity of the trace, we have
\bea &~&\mathcal{F}^{(u)}(L_{-1})-\mathcal{F}^{(d)}(L_{-1})
=\Tr q^{L_0}P_{p}[L_{-1},\phi_1(w_1)\phi_2(w_2)\ldots]=
\sum_{j=1}^n
e^{+iw_j}\left(-h_j+i\frac{\partial}{\partial w_j}\right)
\mathcal{F}_p \notag \\
&~&\mathcal{F}^{(u)}(L_1)-\mathcal{F}^{(d)}(L_1)
=\Tr q^{L_0}P_{p}[L_{+1},\phi_1(w_1)\phi_2(w_2)\ldots]
=\sum_{j=1}^n e^{-iw_j}\left(h_j+i\frac{\partial}{\partial w_j}\right)
\mathcal{F}_p. \notag \\
\eea
Thus, we obtain
\bea &~& \frac{\mathcal{F}^{(u)}(L_0)}{\mathcal{F}_p}=q\frac{\partial}{\partial q}\log \mathcal{F}_p \notag \\
&~&\frac{\mathcal{F}^{(u)}(L_{-1})}{\mathcal{F}_p}=
\frac{q}{1-q}\sum_{j=1}^n e^{iw_j}\left[h_j-i\frac{\partial}{\partial w_j}\log\mathcal{F}_p\right] \notag \\
&~&\frac{\mathcal{F}^{(u)}(L_{1})}{\mathcal{F}_p}=
\frac{1}{1-q}\sum_{j=1}^n e^{-iw_j}\left[h_j+i\frac{\partial}{\partial w_j}\log\mathcal{F}_p\right].
\eea
In the heavy limit, the Casimir equation reads
\begin{equation} \frac{\mathcal{F}^{(u)}(L_1)}{\mathcal{F}_p}\frac{\mathcal{F}^{(u)}(L_{-1})}{\mathcal{F}_p}
-\left(\frac{\mathcal{F}^{(u)}(L_0)}{\mathcal{F}_p}\right)^2=-h_p^2 \,. \end{equation}

\subsubsection{AdS side}

In this section we write AdS in the global coordinates
\begin{equation}
ds^2 = d\rho^2 + \cosh^2 \rho dt^2 + \sinh^2 \rho d\phi^2
\end{equation}
and write $w=\phi+it$. The \sltwo\ Killing vectors are
\bea
\Lc_0 &=& -i \p_w \cr
\Lc_{-1} & = & -i e^{iw} \left({\cosh 2\rho \over \sinh 2\rho}\p_w -{1\over \sinh 2\rho} \p_{\wb} -{i\over 2} \p_\rho\right)  \cr
\Lc_{1} & = & -i e^{-iw} \left({\cosh 2\rho \over \sinh 2\rho}\p_w -{1\over \sinh 2\rho} \p_{\wb} +{i\over 2} \p_\rho\right)
\eea

The torus block $\mathcal{F}_p(\{x_i\},q)$ will be related to the action of a geodesic network,  as in Eq.~(\ref{onshell1}). Most of the steps are similar to that in Sec.~\ref{sec:adsnetwork}, so we shall be terse and only highlight the differences in some of the intermediate steps as well as a few new ingredients in the computations. We also follow the notations in Sec.~\ref{sec:adsnetwork}.

First, let us write the identification that  defines the boundary torus as $w\cong w+ 2\pi  \cong w+\dw$, so that
 $q=e^{2\pi i \tau}$.   We then have
\begin{equation} q\frac{\partial}{\partial q}S_{\text{on-shell}}
(\{x_i\},q)
=\frac{1}{ i}\frac{\partial}{\partial (\dw)} S_{\text{on-shell}}(\{x_i\},q)
.\end{equation}
Focusing on the geodesic that winds around the $\tau$ cycle, roughly speaking, when we increase $\tau$ we are effectively adding in an extra segment of this geodesic, whose unit tangent vector is $\hat{l}_p$.
More concretely, the variation of the on-shell action with respect to $\dw$ can be thought of as a variation of the geodesic action with respect to its endpoints.   As usual, such a variation is given by the canonical momentum conjugate to the displaced coordinate.   The canonical momentum conjugate to $w$-translations, which are isometries generated by $i {\cal L}_0$, is $2ih_p \langle {\cal L}_0,\hat{l}_p \rangle$,\footnote{The calculation is basically the same as that in footnote \ref{fn1}.} and so
\begin{equation}
q\frac{\partial}{\partial q}S_{\text{on-shell}}(\{x_i\},q)
=2h_p \langle {\cal{L}}_0,\hat{l}_p\rangle(v_p)
. \end{equation}
Next, we displace all of the boundary points along a Killing vector $K$ to obtain
\begin{equation}
 \sum_{i=1}^j{\cal{K}}^{(i)} S_{\text{on-shell}}(\{x_i\},q)
 =\sum_{i=1}^j 2h_i \langle {\cal{K}},\hat{l}_i\rangle(x_i).
\end{equation}
As in the sphere $n$-point block case, these geodesics fuse with each other. However, in this case they eventually fuse into the two geodesic end points  which connect with each other around the torus. Thus,
\begin{equation} \sum_{i=1}^j{\cal{K}}^{(i)} S_{\text{on-shell}}(\{x_i\},q)
=2h_p\left[
\langle {\cal{K}}(w_p),\hat{l}_p(w_p+\dw)\rangle
-\langle {\cal{K}}(w_p),\hat{l}_p(w_p) \rangle
\right] \,,\label{aa}
\end{equation}  where $w_p$ denotes the $w$'s coordinates of the bulk vertex $v_p$.

Note that the $h_p$ geodesic has a kink at the location of the vertex, due to the pulling from the other geodesic segment.  Therefore $\hat{l}_p(w_p+\dw) \neq \hat{l}_p(w_p)$ and the two terms on the right hand side do not cancel.  On the other hand, since $\langle {\cal{K}}(w_p+\dw),\hat{l}_p(w_p+\dw)\rangle=\langle {\cal{K}}(w_p),\hat{l}_p(w_p)\rangle$, we can rewrite (\ref{aa}) in terms of the discontinuity of the Killing vector  around the circle, $\delta K = K(w_p+\dw) - K(w_p)$.

Using the explicit form of the Killing vectors now gives
\bea &~& \sum_{k=1}^j-i\frac{\partial}{\partial w_k}S_\text{on-shell}
=-2h_p\langle {\delta \cal{L}}_0,\hat{l}_p\rangle \notag \\
&~& \sum_{k=1}^j-ie^{iw_k}\left(\frac{\cosh 2\rho_k^{(\inf)}}{\sinh 2\rho_k^{(\inf)}}\frac{\partial}{\partial w_k}
-\frac{1}{\sinh 2\rho_k^{(\inf)}}\frac{\partial}{\partial \bar{w}_k}-\frac{i}{2}\frac{\partial}{\partial \rho_k^{(\inf)}}\right)S_\text{on-shell}
=-2h_p\langle {\delta \cal{L}}_{-1},\hat{l}_p\rangle  \notag \\
&~& \sum_{k=1}^j -ie^{-iw_k}\left(\frac{\cosh 2\rho_k^{(\inf)}}{\sinh 2\rho_k^{(\inf)}}\frac{\partial}{\partial w_k}
-\frac{1}{\sinh 2\rho_k^{(\inf)}}\frac{\partial}{\partial \bar{w}_k}+\frac{i}{2}\frac{\partial}{\partial \rho_k^{(\inf)}}\right)S_\text{on-shell}
=-2h_p\langle {\delta \cal{L}}_{1},\hat{l}_p\rangle \notag. \\
\eea
The action of a geodesic approaching the boundary diverges with the cutoff as $S \sim 2h \rho^{(\inf)}$, so
\bea &~& \sum_{k=1}^j-i\frac{\partial}{\partial w_k}S_\text{on-shell}
=-2h_p\langle {\delta \cal{L}}_0,\hat{l}_p\rangle \notag \\
&~& \sum_{k=1}^j -ie^{iw_k}\left(\frac{\partial}{\partial w_k}
S_\text{on-shell}-ih_k \right)
=-2h_p\langle {\delta \cal{L}}_{-1},\hat{l}_p\rangle  \notag \\
&~& \sum_{k=1}^j -ie^{-iw_k}\left(\frac{\partial}{\partial w_k}
S_\text{on-shell}+ih_k\right)
=-2h_p\langle {\delta \cal{L}}_{1},\hat{l}_p\rangle  \notag \\
\eea
The Killing vectors obey
\bea &~&{\cal{L}}_0(w+\dw)={\cal{L}}_0(w) \notag \\
&~&{\cal{L}}_{-1}(w+\dw)=q{\cal{L}}_{-1}(w) \notag \\
&~&{\cal{L}}_1(w+\dw)=\frac{1}{q}{\cal{L}}_1(w), \eea
which leads to
\bea &~&\sum_{k=1}^j \frac{\partial}{\partial w_k}S_\text{on-shell}=0 \notag \\
&~& -2h_p\langle {\cal{L}}_{-1},\hat{l}_p\rangle=\frac{q}{1-q}\sum_{k=1}^j\left(h_k+i\frac{\partial}{\partial w_k}S_\text{on-shell}\right) \notag \\
&~& -2h_p \langle {\cal{L}}_1,\hat{l}_p\rangle=\frac{1}{1-q}\sum_{k=1}^j
\left(h_k-i\frac{\partial}{\partial w_k}S_\text{on-shell}\right).
\eea
Using  the completeness relation
\begin{equation} (2h_p\langle {\cal{L}}_1,\hat{l}_p\rangle)(2h_p\langle{\cal{L}}_{-1},\hat{l}_p\rangle)
-(2h_p\langle {\cal{L}}_0,\hat{l}_p\rangle)^2=-h_p^2\langle \hat{l}_p,\hat{l}_p\rangle=-h_p^2, \end{equation}
we then arrive at the same Casimir equation as on the CFT side, with  the identifications
\bea
&~& \log\mathcal{F}=-S_\text{on-shell}+{\text{constant}} \notag \\
&~& \frac{\mathcal{F}(L_j)}{\mathcal{F}}=-2h_p\langle {\cal{L}}_j,\hat{l}_p\rangle. \eea
This establishes our bulk geodesic description of the heavy torus multi-point block in the semi-classical limit.

\section{Wilson line formulation of conformal blocks}

In three bulk dimensions, there exists an alternative holographic description of conformal blocks based on the Chern-Simons description of 3D gravity. A background metric solving Einstein's equations with negative cosmological constant is described (in Euclidean signature) by a flat $\lie{sl}(2,\CC)$ connection\footnote{Gravity in Lorentzian signature is recovered by an analogous construction with an $\lie{sl}(2,\RR)\oplus \lie{sl}(2,\RR)$ connection.}, so the only gauge-invariant quantities are built from Wilson lines carrying some representation, joined at junctions with a singlet state to maintain gauge invariance, or ending at the boundary where the boundary conditions pick out a preferred gauge. As shown in previous work\cite{Bhatta:2016hpz,Besken:2016ooo}, such networks correspond to conformal blocks, where the representations carried by the Wilson lines correspond to the conformal family of the operator under consideration. In this section, we show that, in general, the Wilson line networks satisfy the Casimir equations of the corresponding conformal blocks, which could be used as an alternative derivation of earlier results. This will include networks in the thermal AdS background, which has the novelty of a non-contractible cycle around which the gauge field has nontrivial holonomy. Including Wilson lines which wrap the thermal cycle, we recover the expected thermal conformal blocks.

\subsection{Chern-Simons gravity}
We briefly collect the required background for convenience and to fix conventions. For a more extensive review, see \cite{Ammon:2012wc}.

In three dimensions, a metric and metric-compatible connection can be defined by a dreibein $e$ and spin-connection $\omega$, both being one-forms valued in the the $\lie{su}(2)$ Lie algebra of $2\times 2$ anti-Hermitian matrices (isomorphic to $\lie{so}(3)$, appearing as the local Lorentz group). Concretely, the metric is given by
\begin{equation}\label{dreibeinMetric}
	g_{\mu\nu}=-2\Tr(e_\mu e_\nu)
\end{equation}
where the trace is taken in the two-dimensional fundamental representation. This metric is automatically covariantly constant under the connection $\omega$, by the Lie algebra invariance property of the quadratic form given by the trace. Now, if we combine the dreibein and connection into the $SL(2,\CC)$ connection $A=\omega + i e$, the flatness of $A$ is equivalent to Einstein's equations with negative cosmological constant (including $\omega$ being torsion-free, so it is the usual Levi-Civita connection). The $SL(2,\CC)$ gauge transformations, decomposed into Hermitian and anti-Hermitian parts, act as the local Lorentz group and as diffeomorphisms (on-shell).

The Einstein-Hilbert action in first-order formalism in terms of these variables becomes a Chern-Simons action, with level determined by Newton's constant and the AdS radius (or by the central charge in the language of the dual CFT). Reproducing the global conformal blocks requires only quantum field theory in a fixed background, without dynamical gravity.  We will therefore focus only on the relationship between flat connections and asymptotically AdS geometries.

Solutions obeying the appropriate boundary conditions, choosing a flat boundary metric written in holomorphic coordinates ($ds^2=dzd\bar{z}$), can be written as the gauge transformation of the manifestly flat connection $a=a(z)dz$:
\begin{equation}\label{CSbackground}
	A=b^{-1} a b + b^{-1}db,\;\text{ with } b=e^{\rho L_0}~,\quad
	a=\left(L_1-2\pi \frac{6 T(z)}{c} L_{-}\right)dz~.
\end{equation}
Here we have chosen a basis for $\lie{sl}(2,\CC)$ spanned by $L_{\pm 1}$ with Lie brackets
\begin{equation}
	[L_{\pm 1},L_0]=\pm L_{\pm 1},\quad [L_1,L_{-1}]=2L_0,
\end{equation}
which may be written in the fundamental representation as
\begin{equation}
	L_0=\frac{1}{2}\begin{pmatrix}1&0\\0&-1\end{pmatrix},\quad L_1=\begin{pmatrix}0&0\\-1&0\end{pmatrix},\quad L_{-1}=\begin{pmatrix}0&1\\0&0\end{pmatrix}.
\end{equation}

Now, computing the metric from the dreibein $e=\tfrac{1}{2i}(A+A^\dagger)$ and \cref{dreibeinMetric}, it is an asymptotically AdS$_3$ metric in the Fefferman-Graham gauge, with the desired boundary metric, and stress-tensor expectation value $T(z)$:
\begin{equation}
	ds^2 = d\rho^2 +e^{2\rho} dz d\bar{z} + 2\pi \frac{6 T}{c} dz^2 + 2\pi \frac{6 \bar{T}}{c} d\bar{z}^2 + e^{-2\rho}\left(\frac{12\pi}{c}\right)^2 T \bar{T} dz d\bar{z}~.
\end{equation}
The boundary metric $ds^2=dzd\bar{z}$ is read off from the leading order piece as $\rho\to\infty$, and the stress tensor from the subleading piece \cite{Balasubramanian:1999re}.

This metric may have singularities somewhere in the bulk (the metric in $(z,\bar{z},\rho)$ coordinates is singular on the surface $e^{2\rho} = \frac{12\pi}{c}|T|$). For our purposes, we want to find the solution corresponding to global AdS$_3$, for which the $z$ coordinate is periodically identified as $z=\phi+ i t \sim z+2\pi$ so the boundary spacetime is a cylinder. The bulk is a solid cylinder in which the spatial circle is contractible, which implies that the holonomy of the gauge field around that cycle must be trivial. This is satisfied by constant stress tensor expectation value $T(z)=-\frac{c}{48\pi}$, so that
\begin{equation}
	a= \left(L_1 +\frac{1}{4}L_{-1}\right)dz
\end{equation}
and this indeed reproduces the usual global AdS$_3$ metric, with $r=\sinh(\rho+\log 2)$.
The zero mode of $T(z)$ is $L_0=\int_0^{2\pi} T(z)dz =-\frac{c}{24}$, corresponding to the usual Casimir energy of the CFT on a circle. Thermal AdS is just this metric with the additional identification $z\sim z+2\pi\tau$. The Euclidean BTZ black hole is constructed similarly by instead trivializing the holonomy around the time circle.

\subsection{The proposal}

Given the ingredients involved in a conformal block (the global conformal representations of the involved operators, and for us the thermal background), there is a simple, natural candidate to construct it from the Chern-Simons formalism. Firstly, the canonical gauge chosen for the background gauge field \cref{CSbackground} is holomorphic, so it is natural to expect gauge invariant constructions involving $a$ to capture the holomorphic piece, and the conjugate connection to pick out the antiholomorphic part. This is special to CFT in two dimensions, where the conformal group factorizes. Having made this comment, we now focus exclusively on the holomorphic sector.

Essentially the only object from which gauge invariant quantities may be constructed is a Wilson line, or holonomy of the gauge field:
\begin{equation}
	W_\alpha[x_0,x_1] = \mathcal{P}\exp\left(-\int_{x_0}^{x_1} a\right)~.
\end{equation}
The subscript $\alpha$ labels a representation $R_\alpha$ of the gauge group, so the holonomy acts to map $R_\alpha$ at the inital point $x_0$ to the end point $x_1$ in a covariant way, transforming as $W_\alpha[x_0,x_1]\rightarrow g_\alpha(x_1) W_\alpha[x_0,x_1] g_\alpha^{-1}(x_0)$ under gauge transformations. The flatness of the connection means that the Wilson line depends only on its endpoints and topology, and not on the details of the path. The Wilson lines may then be connected together into a network, joined at vertices with appropriate singlet states in the tensor products of representations to retain gauge invariance. It will be sufficient to consider only trivalent vertices, and instead of joining using a singlet state in the tensor product of three representations, it will be convenient (and equivalent) to use an intertwining operator $\intertwine{\alpha}{\beta}{\gamma}:R_\alpha\otimes R_\beta\to R_\gamma$, defined to satisfy an invariance property
\begin{equation}
	g_\alpha \intertwine{\alpha}{\beta}{\gamma} g_\beta g_\gamma = \intertwine{\alpha}{\beta}{\gamma}.
\end{equation}
For the irreducible lowest weight representations of $\lie{sl}(2)$ of interest to us, an intertwiner operator is unique (up to normalization) if it exists (since $R_\gamma$ appears at most once in the decomposition of $R_\alpha\otimes R_\beta$).

Finally, we only require gauge invariance under gauge transformations that vanish on the boundary, with the large gauge transformations, which do not vanish at the boundary but preserve the boundary conditions, corresponding to the local conformal group of the CFT. This means that we may end Wilson lines at the boundary, and contract with some canonically chosen state in the relevant representation, for which a natural choice is the lowest weight state, with the smallest eigenvalue of $L_0$ and annihilated by $L_1$, which we will denote $\lw{\alpha}\in R_\alpha$.

The result is a network of Wilson lines $W_\alpha$ in the bulk, carrying specified representations, joined by intertwiners $\intertwine{\alpha}{\beta}{\gamma}$ at trivalent vertices, and ending at the boundary where they are contracted with $\lw{\alpha}$. We will show that this evaluates to a global conformal block, with the endpoints of Wilson lines on the boundary corresponding to external operators, and the internal representations corresponding to the exchanged operators appearing either in the OPE, or for a Wilson line traversing the thermal cycle, a conformal family appearing in the Boltzmann sum. The simplest example is the $\lie{sl}(2)$ character of the representation, which counts the contribution of a global conformal family to the partition function, computed from a Wilson loop round the thermal circle:
\begin{equation}
	\Tr_\alpha\left(W_\alpha[z,z+2\pi\tau]\right)=\chi_\alpha(q)~.
\end{equation}
 Including a trivalent vertex on the Wilson loop and a Wilson line from this vertex to the boundary gives a one-point thermal block:
\begin{equation}
	\Tr_\alpha\left(W_\alpha[z_b,z_b+2\pi\tau](\intertwine{\alpha}{\alpha}{\beta}W_\beta[z_b,z]\lw{\beta}) \right)=\mathcal{F}(\alpha,\beta;q)~.
\end{equation}

\subsection{The Casimir equation from Wilson lines}

The main piece of our argument will be to show that the Wilson line networks satisfy algebraic relations that are precisely analogous to the corresponding objects in the CFT. With these in place, it will follow immediately that the networks satisfy the same Casimir equations as the blocks, since the arguments from CFT will go through unchanged.

The first ingredients in the Wilson line networks are the internal Wilson lines. By the flatness of the connection, we may take all bulk vertices to lie at any point we wish, so in particular they may all be coincident at $z=0$ (the radial position is irrelevant in the gauges we work in). This makes the internal bulk Wilson lines trivial, with the exception of loops with nontrivial topology, wrapping the thermal cycle. These produce factors $\exp\left(-2\pi\tau \left(L_1 +\frac{1}{4}L_{-1}\right)\right)$ in the appropriate representation, which should be analogous to the insertion of $q^{L_0}$ producing the Boltzmann factors in the CFT. As it stands, this is unclear, so it will be helpful to do a constant gauge transformation (or equivalently, a change of canonical $\lie{sl}(2)$ basis) so that the connection is given by $=-i L_0 dz$:
\begin{equation}
	a= \left(L_1+\tfrac{1}{4}L_{-1}\right)dz = g \left(-i L_0 dz\right) g^{-1};\quad g=e^{\frac{i}{2}L_{-1}}e^{-iL_1}e^{-\frac{i\pi}{2}L_0}~.
\end{equation}
The final factor is not required here, but is chosen for later convenience. In this gauge, a Wilson loop traversing the thermal circle is precisely the operator $q^{L_0}$ with $q=e^{2\pi i \tau}$, in the appropriate representation:
\begin{equation}
	W_\alpha[0,2\pi\tau] = e^{2\pi i \tau L_0^\alpha}=q^{L_0^\alpha}~.
\end{equation}

The other ingredient required is the Wilson line running to the boundary, analogous to an insertion of an external operator. With this in mind, we define the operator
\begin{equation}
	[\phi_\gamma(z)]_{\alpha\beta} := \intertwine{\alpha}{\beta}{\gamma}W_\gamma[0,z] \widetilde{\lw{\gamma}}
\end{equation}
from $R_\beta$ to $R_\alpha$. We will show that it satisfies (in the new gauge) an identity, interpreted as the operator transforming as a primary field
\begin{equation}\label{phidentity}
	L_n^\alpha [\phi_\gamma(z)]_{\alpha\beta}-[\phi_\gamma(z)]_{\alpha\beta}L_n^\beta = -\mathcal{L}^\gamma_n[\phi_\gamma(z)]_{\alpha\beta}
\end{equation}
for $n=0,\pm 1$, where $\mathcal{L}^\gamma_n=e^{inz}(i\partial_z-nh_\gamma)$ is the usual differential operator acting on the coordinate $z$, $h_\gamma$ is the lowest weight of the representation $R_\gamma$, and the superscripts on the $L_n$'s indicate the representations in which they are to be taken. The tilde over the lowest weight state is to indicate that after the gauge transformation, it is no longer lowest weight, but has been acted on by $g_\gamma$: $\lw{\gamma}=g_\gamma\widetilde{\lw{\gamma}}$.

After using the infinitesimal version of the invariance property of the intertwiner to pull $L_n^\alpha$ past $\intertwine{\alpha}{\beta}{\gamma}$, to prove the claimed identity it is sufficient to show that $L_n^\gamma+\mathcal{L}_n^\gamma$ annihilates $W_\gamma[0,z]\widetilde{\lw{\gamma}}$. Dropping the $\gamma$ labels, we have
\begin{align}
	(L_n+\mathcal{L}_n)W[0,z]\widetilde{\lw{}} &= (L_n+e^{inz}(i\partial_z-nh))e^{izL_0}\widetilde{\lw{}}\nonumber \\
	&=e^{iz (L_0+n)}(L_n-L_0-nh)\widetilde{\lw{}}
\end{align}
which automatically vanishes for $n=0$, and, taking the sum and difference for $n=\pm 1$, we require that $\widetilde{\lw{}}$ is annihilated by $L_1+L_{-1}-2L_0$, and is an eigenstate of $\frac{1}{2}(L_1-L_{-1})$ with eigenvalue $h$. Now, to see what this implies for the untilded state in the original gauge, undoing the gauge transformation gives $g(L_1+L_{-1}-2L_0)g^{-1}=4iL_1$ and $g\frac{1}{2}(L_1-L_{-1})g^{-1}=L_0$, so the conditions are satisfied precisely when $L_1\lw{}=0$ and $L_0\lw{}=h\lw{}$, so that $\lw{}$ is a lowest weight state of weight $h$.

Now we have the Wilson lops round the thermal circle represented as $q^{L_0}$ by choice of gauge, and external Wilson lines joined to the network by the operators $[\phi_\gamma(z)]_{\alpha\beta}$ satisfying the identity \cref{phidentity}, which is enough to replicate the arguments leading to the Casimir equations like \cref{onepointeq} derived in earlier sections. In that instance, the Casimir evaluated to a constant because of the insertion of projection operators, but here we need no projection, since the operators in all cases are in some definite representation, which encodes the choice of internal conformal multiplets.

\subsection{An explicit example: the characters}

The arguments above show rather abstractly that the Wilson line networks obey the expected Casimir equations of global conformal blocks, which when supplemented with appropriate boundary conditions, is enough to show their equality. In this section, we will make this more concrete in an example, to indicate how direct calculations of the Wilson line networks proceed. We will focus on the simplest case of characters of $\lie{sl}(2,\RR)$, since it is indicative of the sort of combinatorial arguments involved.

From the general proposal above, the character (contribution of a quasiprimary and its global descendants to the partition function) should equal the trace of the holonomy $\gamma=\mathcal{P}\exp\left(-\oint a\right)$ round the thermal circle, in the appropriate representation. Here, we will take the finite dimensional highest weight (non-unitary) representations corresponding to the degenerate operator of weight $h=-n/2$. This is the  $(n+1)$-dimensional representation of $\lie{sl}(2,\RR)$ constructed from the symmetrized tensor product of $n$ fundamental representations. The trace of $\gamma$ in this representation can be written as
\begin{equation}
	\chi_n=\gamma^{(i_1}_{\phantom{i_1}(i_1}\gamma^{i_2}_{\phantom{i_2}i_2} \cdots \gamma^{i_n)}_{\phantom{i_n}i_n)}
\end{equation}
where $\gamma^i_{\phantom{i}j}$ are the matrix elements of $\gamma$ in the fundamental two-dimensional representation. The brackets indicate symmetrization, summing over all permutations and including a factor of $1/n!$.

In this sum over the $n!$ permutations of the $n$ indices $i_1,\ldots,i_n$, each permutation gives a product of traces of powers depending on its cycle structure. We can split up the sum over $S_n$ based on the length $k$ of the cycle containing $1$:
\begin{align}
	\chi_n &= \frac{1}{n!}\sum_{\sigma\in S_n}\gamma^{i_1}_{\phantom{i_1}i_{\sigma(1)}}\gamma^{i_2}_{\phantom{i_2}i_{\sigma(2)}} \cdots \gamma^{i_n}_{\phantom{i_n}i_{\sigma(n)}}\\
	&= \frac{1}{n!}\sum_{k=1}^n (n-1)!\Tr(\gamma^k) \chi_{n-k}~.
\end{align}
Here the factor $(n-1)!$ is the number of permutations with $1$ in a cycle of length $k$ (which turns out to be independent of $k$), $\Tr(\gamma^k)$ is the contribution from the cycle containing $1$, and $\chi_{n-k}$ accounts for the permutations of the remaining indices. From $\Tr(\gamma^k)=q^{n/2}+q^{-n/2}$ (since $q^{\pm\frac{1}{2}}$ are the eigenvalues of $\gamma$), and the `initial condition' $\chi_0=1$, this recursively computes the characters for all $n$, giving
\begin{equation}
	\chi_n(q) = q^{-\frac{n}{2}}\frac{1-q^{n+1}}{1-q},
\end{equation}
which can be proved by induction on $n$. This is the expected answer, counting one state at each level between $-\frac{n}{2}$ and $\frac{n}{2}$, with higher weights being annihilated since this is a degenerate representation.

\section*{Acknowledgements}

  We would like to thanks James Sully for helpful discussions. P.K. is supported in part by NSF grant PHY-1313986.  A. M. and H. M are supported by the National Science and Engineering Council of Canada and by the Simons Foundation.  G. N. is supported by the National Science and Engineering Council of Canada. J.-q.W. is supported in part by NSFC Grant No. 11275010, No. 11335012 and No. 11325522.

\bibliographystyle{ssg}
\bibliography{biblio}

\end{document}